\documentclass[preprint]{ptephy_v1}%%%%%% to generate preprint number
\usepackage{amsmath,amssymb}
\usepackage{color}
\usepackage{mathrsfs}
\usepackage{subcaption}

\preprintnumber{CHIBA-EP-253v2} %%% %%% Insert preprint number here

\usepackage{bm}

\usepackage{braket}

\allowdisplaybreaks

\begin{document}

\title{Color confinement and restoration of residual local gauge symmetries}

\author[1]{Kei-Ichi Kondo}
\affil{Department of Physics, Graduate School of Science, Chiba University, Chiba 263-8522, Japan\email{kondok@faculty.chiba-u.jp}}

\author{Naoki Fukushima}
\affil{Department of Physics, Graduate School of Science and Engineering, Chiba University, Chiba 263-8522, Japan\email{segapsn913@gmail.com}}

%%% To include the collaborator name... Please use the command "\collaborator"
%%% For example: \collaborator{ATLAS Collaboration}

\begin{abstract}%
All colored particles including dynamical quarks and gluons are confined if the color confinement criterion proposed by Kugo and Ojima is satisfied. 
The criterion was obtained under the gauge fixing of the Lorenz type. 
However, it was pointed out that the Kugo-Ojima criterion breaks down for the Maximal Abelian gauge, which is quite strange in view of the fact that quark confinement has been verified according to the dual superconductivity caused by magnetic monopole condensations.  
In order to make a bridge between color confinement due to Kugo and Ojima and the dual superconductor picture for quark confinement, we investigate a generalization of the color confinement criterion to obtain the unified  picture for confinement.
We show that the restoration of the residual local gauge symmetry which was shown in the Lorenz gauge by Hata to be equivalent to the Kugo-Ojima criterion  indeed occurs  in the Maximal Abelian gauge for the SU(N) Yang-Mills theory in two-, three- and four-dimensional Euclidean spacetime once the singular topological configurations of gauge fields are taken into account. 
This result indicates that the color confinement phase is a disordered phase caused by non-trivial topological configurations irrespective of the gauge choice. 
As a byproduct, we show that the compact U(1) gauge theory can have the disordered confinement phase, while the non-compact U(1) gauge theory has the deconfined Coulomb phase.

\end{abstract}

\subjectindex{...}

\maketitle

%\newpage

\noindent
%\textit{1. Introduction.} \ 
\section{Introduction}

Color confinement is still an unsolved problem in particle physics, although it is expected that a deep understanding of the quantum gauge field theory will enable us to explain why color confinement really occurs due to strong interactions.
For quark confinement, especially, the dual superconductor picture has succeeded to give a good answer to the question as to why quark confinement occurs, although more than 40 years have passed after the original proposal \cite{dualsuper}. 
See e.g., \cite{CP97} for a review in the early stage and \cite{KKSS15} for a review of recent developments on quark confinement from the viewpoint of  the dual superconductivity which is caused by condensations of magnetic monopoles  in the gauge-independent manner. 
Even if the dual superconductor picture is true,  however, it is not an easy task to apply this picture to various composite particles composed of quarks and/or gluons. 
In fact,  gluon confinement is still less understood, although there are interesting developments quite recently, see \cite{HK21} and reference therein. 

In view of these, we recall the color confinement criterion derived by Kugo and Ojima long ago \cite{KO79}. 
The Kugo--Ojima (KO) criterion was given within the manifestly Lorentz covariant operator formalism on an indefinite metric state space $\mathcal{V}$ by assuming the existence of the nilpotent BRST symmetry under the Lorentz covariant gauge fixing condition.  
If the KO criterion is satisfied, all colored objects cannot be observed.  
Therefore, the KO color confinement is a very strong and attractive statement because quark confinement and gluon confinement immediately follow as special cases of color confinement once color confinement in the sense of Kugo and Ojima was proved \cite{Kugo89,Kugo95,Hata82,Hata83}.

However, the KO criterion was derived only in the Lorenz gauge $\partial^\mu \mathscr{A}_\mu=0$, even if the issue on the existence of the nilpotent BRST symmetry is put aside for a while. 
The KO criterion is written in terms of a specific correlation function called the KO function which is clearly gauge-dependent and is not directly applied to the other gauge fixing conditions. 
Color confinement regarded as a physical phenomenon must be understood in the gauge-independent manner. 
However, such a gauge-independent criterion for color confinement has never been found to the best of the author knowledge, in sharp contrast to quark confinement for which the gauge-invariant Wilson criterion is as well-known written in terms of the vacuum expectation value of the Wilson loop operator  \cite{Wilson74}. 
At this moment, therefore, it will be the first step to find the color confinement criterion comparable to the KO criterion in the gauges other than the Lorenz gauge to consider the physical meaning of the KO criterion. 

From this point of view, the maximally Abelian (MA) gauge \cite{MAG,KondoII,Kondo01} is the best gauge to be investigated  because the dual superconductor picture for quark confinement was intensively investigated in the MA gauge for which  magnetic monopoles can be defined even in the pure Yang-Mills theory without any scalar field and the condensation of magnetic monopoles is shown to take  place to yield quark confinement according to the Wilson criterion at least in numerical simulations. 
Nevertheless, it has been pointed out by Suzuki and Shimada \cite{SS83} that the KO criterion cannot be applied to the MA gauge and the KO criterion is violated in the model for which quark confinement is shown to occur \cite{Polyakov77} due to magnetic monopole and antimonopole condensation in accord with the dual superconductor picture. 
Subsequently, it was claimed by Hata and Niigata \cite{HN93} that the MA gauge is an exceptional case to which the KO color confinement criterion cannot be applied. 
If these claims were true, we would have no theoretical method to characterize color confinement. 
We wonder how the color confinement criterion of the KO type is compatible with the dual superconductor picture for quark confinement.

The purpose of this paper is to revisit this issue, namely, gauge-independence of color confinement criterion of the KO type.
Of course, the explicit expression of such a criterion of the KO type, even if it exists, depends on the gauge choice and changes its form gauge by gauge. 
However, the same idea as the KO in the Lorenz gauge must be applied to any gauge without failure. 
In this paper we reconsider the color confinement criterion of the KO type in the Lorenz gauge and give an explicit form to be satisfied in the MA gauge within the same framework as the Lorenz gauge in the manifestly Lorentz covariant operator formalism with the unbroken BRST symmetry. 
Moreover, we show that the criterion is indeed satisfied in the MA gauge, once the effect of the topological configurations are taken into account in the criterion. 
If we ignore such an effect, the criterion is not satisfied in agreement with the previous investigations \cite{SS83,HN93}. 
In other words, such topological effect is essential to achieve color confinement compatible with the dual superconductor picture caused by condensation of topological objects.
Therefore, our results in this paper give a bridge between the color confinement of the KO type and the dual superconductor picture for quark confinement. 

For this purpose, we make use of the method of Hata \cite{Hata82} saying that 
the KO criterion is equivalent to the condition for the residual local gauge symmetry to be restored. 
We show that singular topological gauge field configurations play the role of restoring the residual local gauge symmetry violated in the MA gauge.
This result implies that color confinement phase is a disordered phase which is realized by non-perturbative effect due to topological configurations. 
As a byproduct, we show that the Abelian U(1) gauge theory in the compact formulation can confine electric charges even in $D=4$ spacetime dimensions as discussed long ago by Polyakov \cite{Polyakov75}  in the phase where topological objects recover the residual local gauge symmetry.

This paper is organized as follows.
In section 2, we review the  results obtained by Hata for the KO criterion in the Lorenz gauge. 
In section 3, we give a brief review on deconfinement in Abelian gauge theory in the non-compact formulation. 
In section 4, we examine the restoration of the residual local gauge symmetry in the Lorenz gauge. 
In section 5,  we examine the MA gauge. 
%In section 6, we go back to the Abelian gauge theory in the compact formulation.
The final section is devoted to conclusion and discussion. 
The details of the calculations are given in Appendices.

\section{Color confinement and residual local gauge symmetry restoration}

First of all, we recall the result of Kugo and Ojima on color confinement. 

\noindent
\textbf{Proposition 1}:
[Kugo-Ojima color confinement criterion \cite{KO79}]
In the manifestly Lorentz covariant operator formalism on the indefinite metric state space $\mathcal{V}$, choose the Lorenz gauge fixing $\partial^\mu \mathscr{A}_\mu=0$. 
Suppose that the (nilpotent) BRST symmetry exists.
Let  $\mathcal{V}_{\rm phys}$ be the physical state space with a semi-positive definite metric $\langle {\rm phys} |  {\rm phys}\rangle \ge 0$ as a subspace of  an indefinite metric state space 
$\mathcal{V}$ defined by using the BRST charge operator $Q_{\rm B}$  as
\begin{align}
 \mathcal{V}_{\rm phys} = \{ | {\rm phys}\rangle \in  \mathcal{V};  Q_{\rm B} |{\rm phys}\rangle=0 \} \subset   \mathcal{V} .
\label{KO-cond}
\end{align}
Introduce the function $u^{AB}(p^2)$ defined by
\begin{align}
 u^{AB}(p^2) \left( g_{\mu\nu} - \frac{p_\mu p_\nu}{p^2} \right) =  \int d^Dx \ e^{ip(x-y)} \langle 0 |  {\rm T}^{}[(\mathscr{D}_\mu\mathscr{C})^A (x) g(\mathscr{A}_\mu \times \bar{\mathscr{C}})^B(y) | 0 \rangle,
\label{color-KO-f}
\end{align}
where $\mathscr{C}$ and $\bar{\mathscr{C}}$ are respectively the ghost and antighost fields, and $\mathscr{D}_\mu$ is the covariant derivative in the adjoint representation defined by $\mathscr{D}_\mu \mathscr{C}=\partial_\mu \mathscr{C} -ig [\mathscr{A}_\mu, \mathscr{C}]$. 
If the following condition in the Lorenz gauge  is satisfied
\begin{align}
  \lim_{p^2 \to 0} u^{AB}(p^2) = -\delta^{AB} ,
\label{CCF-KO-c}
\end{align}
then the color charge operator $Q^A$ is well defined, namely, the color symmetry is not spontaneously broken, and 
  $Q^A$ vanishes for any physical state $ \Phi, \Psi \in \mathcal{V}_{\rm phys}$,
\begin{align}
 \langle \Phi | Q^{A} | \Psi \rangle=0 , \quad \Phi, \Psi \in \mathcal{V}_{\rm phys} .
\label{color-KO2}
\end{align}
%provided that the remnant gauge symmetry with respect to spacetime-independent gauge transformation remains unbroken. 
The BRS singlets as physical particles are all color singlets, while colored particles belong to the BRS quartet representation. 
Therefore, all colored particles cannot be observed and only color singlet particles can be observed.    

The condition (\ref{CCF-KO-c}) is called the \textbf{Kugo-Ojima (KO) color confinement condition}, or \textbf{Kugo-Ojima (KO) color confinement criterion}. 
The function $u^{AB}(p^2)$ is called the \textbf{KO function} and $u^{AB}(0):=\lim_{p^2 \to 0} u^{AB}(p^2)$ is called the \textbf{KO parameter}.

The usual gauge fixing condition is sufficient to fix the gauge in the perturbative framework in the sense that it enables us to perform perturbative calculations.
However, it  does not eliminate the gauge degrees of freedom entirely but leaves certain gauge symmetry which we call the \textbf{residual local gauge symmetry}.  The residual local gauge symmetry can in principle be spontaneously broken.  
This phenomenon does not contradict the Elitzur theorem \cite{Elitzur75}: any local gauge symmetry cannot be spontaneously broken, because the Elitzur theorem does not apply to the residual local gauge symmetries left after the usual gauge fixing. 
The residual symmetries can be both dependent and independent on spacetime coordinates. 

Indeed, Hata \cite{Hata82} investigated the possibility of the restoration of the residual  \textbf{``local gauge symmetry''} in non-Abelian gauge theories with covariant gauge fixing, which is broken in perturbation theory due to the presence of massless gauge bosons even when the \textbf{global gauge symmetry} is unbroken. 
Note that ``local gauge symmetry'' with the quotation marks means that it is not exactly conserved, but is conserved only in the physical subspace $\mathcal{V}_{\rm phys}$ of the state vector space $\mathcal{V}$.
Then he found that this restoration is equivalent to the realization of color confinement due to Kugo and Ojima as summarized below.

%---
%Abstract of Hata82
%Restoration of the local gauge symmetry and its connection to color confinement is investigated in non-Abelian gauge theories with covariant gauge fixing. 
%We consider the Noether current $J^\mu{}\omega$ of the local gauge transformation with transformation functions $\omega(x)$ linear in $x^\mu$.
%This current is conserved only in the physical subspace of the state vector space and in perturbation theory contains a massless pole communicating to the gauge field. 
%We define the local gauge symmetry restoration as the disappearance of this massless "Nambu-Goldstone"pole from $J^\mu{}\omega$. 
%The restoration condition is obtained and it coincides exactly with the color confinement criterion proposed earlier by Kugo and Ojima. 
%Quarks and other colored particles are shown to be confined in the local gauge symmetry restored phase by using the Ward identities of $J^\mu{}\omega$.
%---

\noindent
\textbf{Proposition 2}:
[Hata \cite{Hata82}]
Consider the residual ``local gauge symmetry'' specified by the Lie-algebra-valued transformation function $\omega(x) \in su(N) $ linear in $x^\mu$:
\begin{align}
 \omega(x)  = T_A \omega^A(x) , \  \omega^A(x) = \epsilon_\rho^A x^\rho ,
\label{color-Hata1}
\end{align}
where $\epsilon_\rho^A$ is $x$-independent constant parameters. 
Then there exists the Noether current $\mathscr{J}^\mu{}_\rho^A (x)$  given by
\begin{align}
 \mathscr{J}^\mu_\omega(x)  = gJ^\mu{}^A(x)  x^\rho \epsilon_\rho^A - \mathscr{F}^{\mu\rho}{}^A(x) \epsilon_\rho^A := \mathscr{J}^\mu{}_\rho^A (x) \epsilon^\rho{}^A ,
\label{color-Hata1b}
\end{align}
which is conserved only in the physical subspace $\mathcal{V}_{\rm phys}$ of the state vector space $\mathcal{V}$: 
\begin{align}
 \langle \Phi | \partial_\mu \mathscr{J}^\mu_\omega(x) | \Psi \rangle=0 , \quad \Phi, \Psi \in \mathcal{V}_{\rm phys} ,
\label{color-KO2}
\end{align}
where $J^\mu{}^A(x)$ is the Noether current  associated with the global gauge symmetry which is conserved in the whole state vector space $\mathcal{V}$. 
%For this choice of  the residual ``local gauge symmetry'', 
The Ward-Takahashi (WT) relation holds for the local gauge current $\mathscr{J}^\mu{}_\rho^A (x) $ communicating to the gauge field $\mathscr{A}_\sigma^B(y)$:
\begin{align}
 \int d^Dx \ e^{ip(x-y)} \partial_\mu^x \langle 0 |  {\rm T}^{} [\mathscr{J}^\mu{}_\rho^A (x)  \mathscr{A}_\sigma^B(y)] | 0 \rangle 
 = i \left( g_{\rho\sigma} - \frac{p_\rho p_\sigma}{p^2} \right) [\delta^{AB}+u^{AB}(p^2)] .
\label{color-Hata1}
\end{align}
Thus, if the KO condition  in the Lorenz gauge is satisfied
\begin{align}
  \lim_{p^2 \to 0} u^{AB}(p^2) = -\delta^{AB} ,
\label{color-KO-c2}
\end{align}
then the massless ``Nambu-Goldstone pole'' between the local gauge current  $\mathscr{J}^\mu{}_\rho^A$ and the gauge field $\mathscr{A}_\sigma^B$ contained in perturbation theory disappears.  
It is remarkable that the restoration condition coincides exactly with the Kugo and Ojima color confinement criterion. 
This means that the residual local gauge symmetry is restored if the KO condition is satisfied. 
We define the \textbf{restoration of the residual ``local gauge symmetry''} as the \textbf{disappearance of the massless ``Nambu-Goldstone pole'' } from the local gauge current $\mathscr{J}^\mu{}_\rho^A (x)$ communicating to the gauge field $\mathscr{A}_\sigma^B(y)$ through the WT relation.
In this sense, quarks and other colored particles are shown to be confined in the local gauge symmetry restored phase.

\section{The residual gauge symmetry  in the Abelian gauge theory}

Prior to the analysis of the non-Abelian gauge theory, we give a brief review of the residual local gauge symmetry in the Abelian gauge theory, because understanding the Abelian case is very important to consider the non-Abelian gauge theory, especially in the MA gauge later.

Consider QED, or any local U(1) gauge-invariant system described by the total Lagrangian density
\begin{align}
\mathscr{L} &= \mathscr{L}_{\text{inv}} + \mathscr{L}_{\text{GF+FP}} .
\end{align}
Here $\mathscr{L}_{\text{inv}}$ is a part  of the Lagrangian density  invariant under the local gauge transformation:
\begin{align}
  A_\mu(x) \to A_\mu^\omega(x) := A_\mu(x)+\partial_\mu \omega(x) .
\end{align}
To fix this gauge degrees of freedom, we introduce the Lorenz gauge fixing condition:
\begin{align}
  \partial^\mu A_\mu(x)=0 .
\end{align}
Then the gauge-fixing (GF) and the Faddeev-Popov (FP) ghost term is given by
\begin{align}
  \mathscr{L}_{\text{GF+FP}} 
 = % - \partial_\mu \mathscr{B} \cdot \mathscr{A}^\mu 
  B \partial^\mu A_\mu(x) 
  + \frac{1}{2} \alpha {B}^2 - i \partial^\mu \bar{{c}} \partial_\mu {c} .
%= \bm{\delta} \left( -i \bar c(\partial^\mu A_\mu + \frac12 \alpha B) \right) .
\end{align}

However, this gauge-fixing still leaves the invariance under the transformation function 
 $\omega(x)$ linear in $x^\mu$:
\begin{align}
 \omega(x)  =  a+  \epsilon_\rho x^\rho  ,
\end{align}
since this is a solution of the equation:
%$\delta^\omega A_\mu(x) :=A_\mu^\omega(x) - A_\mu(x)=\partial_\mu \omega(x)=\epsilon_\mu$ and $\delta^\omega \partial_\mu A_\mu(x)=0$.
\begin{align}
\partial^\mu \partial_\mu \omega(x)=0 \Longrightarrow 
   \partial^\mu A_\mu^\omega(x) = \partial^\mu A_\mu(x)+\partial^\mu \partial_\mu \omega(x)  =0  .
\end{align}
This symmetry is an example of the residual local gauge symmetry to be discussed in this paper in detail. 
This fact leads to two conserved charges, the usual charge $Q$ and the vector charge $Q^\mu$, which are generators of the transformation:
\begin{align}
 \delta^\omega A_\mu(x):=A_\mu^\omega(x) - A_\mu(x)
 =  [i(aQ+ \epsilon_\rho Q^\rho  ), A_\mu(x)] 
%\nonumber\\
 =  \partial_\mu \omega(x) = \epsilon_\mu . 
\end{align}
This relation must hold for arbitrary $x$-independent constants $a$ and $\epsilon_\mu$, leading to the commutator relations: 
\begin{align}
  [i Q , A_\mu(x)] = 0, \quad 
[i Q^\rho , A_\mu(x)] = \delta^\rho_\mu . 
\end{align}
The first equation implies that the usual $Q$ symmetry, i.e., the global gauge symmetry is  not spontaneously broken:
\begin{align}
 \langle 0|[i Q  , A_\mu(x)]|0 \rangle = 0 ,
\end{align}
while the second equation implies that $Q^\mu$ symmetry, i.e., the residual local gauge symmetry is  always spontaneously broken:
\begin{align}
 \langle 0|[i Q^\rho , A_\mu(x)]|0 \rangle = \delta_\mu^\rho .
\end{align}
%Thus, $A_\mu(x)$ contains the massless NG boson. 
Ferrari and Picasso \cite{FP71} argued from this observation that photon is understood as the massless Nambu-Goldstone (NG) vector boson associated with the spontaneous breaking of $Q^\mu$ symmetry according to the Nambu-Goldstone theorem. 
However, the above relations alone do not guarantee the existence of massless spin-1 photon. % nor spin-2 graviton. 
Because the spontaneous breaking relations are merely the results of massless poles of  a 2-point function. % as shown below.
For the details, see e.g., \cite{KTU85}.

\begin{comment}
\noindent
\textbf{Proposition 3}:
%\begin{prop}
In the U(1) gauge theory, the massless pole singularities responsible to 
\begin{align}
 \langle 0|[i Q^\rho , A_\mu(x)]|0 \rangle = \delta_\mu^\rho .
 \label{U1-commutator}
\end{align}
are solely given by the 2-point function
\begin{align}
{\rm F.T.} \langle 0 | {\rm T}  {B} (x) {A}_{\mu}(y) | 0 \rangle
 =  \frac{p_\mu}{p^2}  .
 \label{U1-FT}
\end{align}
\end{prop}
\end{comment}
%
\noindent
%[Proof]

The Noether currents associated with the $Q$ and $Q^\mu$ symmetries are respectively given by
\begin{align}
%Q & \leftarrow 
J^\mu =  - \partial_\nu F^{\mu\nu}  - \partial^\mu B , \quad
%\nonumber\\
%Q^\mu & \leftarrow
  J^{\mu\rho} =  B \partial^\mu x^\rho - \partial^\mu B x^\rho - \partial_\nu (x^\rho F^{\mu\nu}) .
\end{align}
%This is valid even for the general U(1) gauge invariant system if the field strength $F^{\mu\nu}$ is replaced by other suitable antisymmetric tensor.
%We can take, as generators by commutator on any local operator
The generators of the transformations are given by
\begin{align}
 Q = \int d^{D-1}x \ J^0(x) , \quad Q^\rho = \int d^{D-1}x \ J^{0\rho}(x)  . 
\end{align}
%We can identify the massless singularity responsible for (\ref{U1-commutator}) with 2-point function $\langle 0 | {\rm T}  B(x) A_{\lambda}(y) | 0 \rangle$ as follows.
The vacuum expectation value (VEV) of the commutator is rewritten as 
\begin{align}
   \langle 0|[i Q^\rho , A_\lambda(y)]|0 \rangle 
%\nonumber\\
 =&  \int d^Dx \ \delta(x^0) \langle 0|[i J^{0\rho}(\bm{x},x^0) , A_\lambda(y)]|0 \rangle 
 \nonumber\\
  =& \lim_{p \to 0} \int d^Dx \ e^{ip(x-y)} i \partial_\mu^{x} \langle 0| {\rm T}^{} J^{\mu\rho}(x) A_\lambda(y) |0 \rangle .
 \end{align}
By inserting  $J^{\mu\rho}$ which has the equivalent form
\begin{align}
 J^{\mu\rho} %= B \partial^\mu x^\rho - x^\rho \partial^\mu B  
 = - \partial^\mu ( x^\rho B) + 2 \eta^{\rho\mu} B - \partial_\nu (x^\rho F^{\mu\nu})  , 
\end{align}
we have 
 \begin{align}
  \langle 0|[i Q^\rho , A_\lambda(y)]|0 \rangle 
%\nonumber\\
  =& \lim_{p \to 0} \int d^Dx \ e^{ip(x-y)} i \partial_\mu^{x} [ - \langle 0| {\rm T}^{} \partial^\mu (x^\rho B(x) ) A_\lambda(y) |0 \rangle 
\nonumber\\    &\quad\quad\quad\quad\quad\quad 
   + 2 \eta^{\rho\mu} \langle 0| {\rm T}^{}  B(x) A_\lambda(y) |0 \rangle ]
 \nonumber\\  
 =& \lim_{p \to 0} \left( p^2 \frac{\partial}{\partial p^\rho} + 2p^\rho \right) \int d^Dx \ e^{ip(x-y)}   \langle 0 | {\rm T}^{}  B(x) A_{\lambda} (y) | 0 \rangle ,
 \end{align}
where we have used the fact that the last term of $J^{\mu\rho}$ does not contribute to the integration. 
The calculation of the VEV is reduced to the 2-point function $\langle 0 | {\rm T}^{}  B(x) A_{\lambda} (y) | 0 \rangle$. 
By using the Fourier transform% of the 2-point function $\langle 0 | {\rm T}^{}  B(x) A_{\lambda} (y) | 0 \rangle$, 
\begin{align}
{\rm F.T.} \langle 0 | {\rm T}^{}  {B} (x) {A}_{\lambda}(y) | 0 \rangle  =  \frac{p_\lambda}{p^2} ,
 \label{U1-FT}
\end{align}
we conclude
\begin{align}
 \langle 0|[i Q^\rho , A_\lambda(y)]|0 \rangle 
%\nonumber\\
 =  \lim_{p \to 0} \left( p^2 \frac{\partial}{\partial p^\rho} + 2p^\rho \right) \frac{p_\lambda}{p^2} 
%\nonumber\\
  =  \lim_{p \to 0} \left[ p^2 \left( \frac{\delta^\rho_\lambda}{p^2}  - p_\lambda \frac{2p^\rho}{p^4}  \right) + 2p^\rho \frac{p_\lambda}{p^2}  \right] 
  = \delta_\lambda^\rho .
 \end{align}
%Using these we can show the proposition in the same way as the tree  level case, although we need a suitable WT identity to prove it at full order level.
%[End of Proof]

Anyway, the restoration of the residual local gauge symmetry does not occur in the ordinary  Abelian case.

%\newpage
%%%%%%%%%%%%%%%%%%%%%%%%%%%%%%%%%%%%%%%%%%%%%%%%%%%%%%%%%
%%%%%%%%%%%%%%%%%%%%%%%%%%%%%%%%%%%%%%%%%%%%%%%%%%%%%%%%%
\section{Residual local gauge symmetry restoration and color confinement in the Lorenz gauge}
%%%%%%%%%%%%%%%%%%%%%%%%%%%%%%%%%%%%%%%%%%%%%%%%%%%%%%%%%
%%%%%%%%%%%%%%%%%%%%%%%%%%%%%%%%%%%%%%%%%%%%%%%%%%%%%%%%%

We judge the \textbf{restoration of the residual local gauge symmetry} by the disappearance of the massless ``Nambu-Goldstone'' pole from the Noether current $\mathscr{J}^\mu_\omega(x)$ associated with the residual ``local gauge symmetry'' communicating to the gauge field $\mathscr{A}_\lambda(y)$. 
We regard this restoration as the color confinement criterion.

The total Lagrangian density is given by
\begin{align}
\mathscr{L} &= \mathscr{L}_{\text{inv}} + \mathscr{L}_{\text{GF+FP}} .
\label{eq:QCD-Lagrangian}
%\mathscr{L}_{\text{inv}} &= - \frac{1}{4} \mathscr{F}_{\mu \nu} \cdot \mathscr{F}^{\mu \nu} + \mathscr{L}_{\text{matter}} (\varphi , D_\mu \varphi) .
\end{align}
The first term $\mathscr{L}_{\text{inv}}$ is the gauge-invariant part for the gauge field $\mathscr{A}_\mu$ and the matter field $\varphi$ given by
\begin{align}
%\mathscr{L} &= \mathscr{L}_{\text{inv}} + \mathscr{L}_{\text{GF+FP}} ,
%\label{eq:QCD-Lagrangian}\\
  \mathscr{L}_{\text{inv}} &= - \frac{1}{4} \mathscr{F}_{\mu \nu} \cdot \mathscr{F}^{\mu \nu} + \mathscr{L}_{\text{matter}} (\psi , D_\mu \psi) ,
\end{align}
with the field strength $\mathscr{F}_{\mu \nu}$ of the gauge field defined by
$
  \mathscr{F}_{\mu \nu} := \partial_\mu \mathscr{A}_\nu - \partial_\nu \mathscr{A}_\mu + g \mathscr{A}_\mu \times \mathscr{A}_\nu = - \mathscr{F}_{\nu \mu} 
%\label{eq:nonabelian-field-strength}
$
and  the covariant derivative $D_\mu \psi$ of the matter field $\psi$ defined by
$D_\mu \psi:=\partial_\mu \psi -ig\mathscr{A}_\mu \psi$.
The second term $\mathscr{L}_{\text{GF+FP}}$  is the sum of the the gauge-fixing (GF) term and the Faddeev-Popov (FP) ghost term 
where the GF term includes the Nakanishi-Lautrup field $\mathscr{B} (x)$ which is the Lagrange multiplier field to incorporate the gauge fixing condition and the FP ghost term includes the ghost field $\mathscr{C}$ and the antighost field $\bar{\mathscr{C}}$. 

For the gauge field and the matter field, we consider  the local gauge transformation specified by the Lie algebra-valued transformation function $\omega (x)=\omega^A(x)T_A$ given by
\begin{align}
  \delta^{\omega} \mathscr{A}_\mu (x) &= \mathscr{D}_\mu \omega (x) := \partial_\mu \omega (x)  + g\mathscr{A}_\mu \times \omega (x) , \nonumber\\
  \delta^\omega \varphi (x) &= i g \omega (x) \varphi (x) .
  \label{eq:nonabelian-gauge-transformation}
\end{align}
In what follows we use the notation
$(\mathscr{A} \times \mathscr{B})^A= f^{ABC} \mathscr{A}^B \mathscr{B}^C$ with the structure constant $f^{ABC}$ for the $SU(N)$ group with the indices $A,B,C =1, \ldots, {\rm dim}SU(N)=N^2-1$. 
For the other fields $\mathscr{B} , \mathscr{C} , \bar{\mathscr{C}}$, the local gauge transformation is just given by localizing the global transformation as
\begin{align}
  \delta^{\omega} \mathscr{B} (x) &= g \mathscr{B} (x) \times \omega (x) , \nonumber\\
  \delta^{\omega} \mathscr{C} (x) &= g \mathscr{C} (x) \times \omega (x) , \nonumber\\
  \delta^{\omega} \bar{\mathscr{C}} (x) &= g \bar{\mathscr{C}} (x) \times \omega (x) .
  \label{eq:gauge-transformation-of-C-B}
\end{align}

Let $\bm{\delta}_{\rm B}$ be the BRST transformation and $Q_B$ be the BRST charge as the generator of the BRST transformation: 
\begin{align}
  &[i Q_B , \mathscr{A}_\mu] = \bm{\delta}_{\rm B} \mathscr{A}_\mu = \mathscr{D}_\mu \mathscr{C} , &&[i Q_B , \mathscr{B}] = \bm{\delta}_{\rm B} \mathscr{B} = 0 , 
  \nonumber\\
  &\{i Q_B , \mathscr{C}\} = \bm{\delta}_{\rm B} \mathscr{C} = - \frac{1}{2} g \mathscr{C} \times \mathscr{C} , &&\{i Q_B , \bar{\mathscr{C}}\} = \bm{\delta}_{\rm B} \bar{\mathscr{C}} = i \mathscr{B} , 
  \nonumber\\
  &[i Q_B , \varphi] = \bm{\delta}_{\rm B} \varphi = i g \mathscr{C} \varphi , &&\{i Q_B , \psi\} = \bm{\delta}_{\rm B} \psi = i g \mathscr{C} \psi .
  \label{eq:BRST-transformation}
\end{align}
The physical state condition is given by 
\begin{align}
  \ket{\text{phys}} \in \mathcal{V}_{\text{phys}} \Leftrightarrow \ Q_B \ket{\text{phys}} = 0 .
  \label{eq:physical-condition}
\end{align}
In particular, the vacuum state $\ket{0}$ satisfies this condition.

Now we proceed to write down the Ward-Takahashi relation to examine the appearance or disappearance of the massless  ``Nambu-Goldstone pole''. 
We consider the condition for the restoration of the residual local gauge symmetry for a general $\omega$. 
We focus on the WT relation 
\begin{align}
 & \int d^D x e^{i p (x - y)} \partial_\mu^x \braket{ {\rm T}^{} \mathscr{J}_\omega^\mu (x) \mathscr{A}_\lambda^B (y)} \nonumber\\
  = & i %\int d^D x \ e^{i p (x - y)} \delta^D (x - y)  
  \braket{\delta^\omega \mathscr{A}_\lambda^B(y)} 
  + \int d^D x \  e^{i p (x - y)} \braket{ {\rm T}^{} \partial_\mu \mathscr{J}_\omega^\mu (x) \mathscr{A}_\lambda^B (y)}   \nonumber\\
%= &i \braket{(\mathscr{D}_\lambda [\mathscr{A}]\omega)^B(y)}  + \int d^D x e^{i p (x - y)} \braket{ {\rm T}^{} \partial^x_\mu {\mathscr{J}_\omega^{\mu }}_\nu^A (x) \partial^\nu \omega^A(x) \mathscr{A}_\lambda^B(y)} \nonumber\\
  = & i \braket{ \partial_\lambda \omega^B(y) + g(\mathscr{A}_\lambda \times \omega)^B(y) } + \int d^D x e^{i p (x - y)} \braket{ {\rm T}^{}\delta^\omega \mathscr{L}_{\rm GF+FP}(x)  \mathscr{A}_\lambda^B(y)} \nonumber\\
  = &  i \partial_\lambda \omega^B (y) + \int d^D x e^{i p (x - y)} \braket{ {\rm T}^{} \delta^\omega \mathscr{L}_{\rm GF+FP}(x)  \mathscr{A}_\lambda^B(y)}    ,
  \label{CCF-WTI_gen1}
\end{align}
where we have assumed the unbroken Lorentz invariance to use $\braket{0 | \mathscr{A}_\lambda (x) | 0} = 0$  in the final step. 
Note that this relation is valid for any choice of the gauge fixing condition. 
See Appendix A for the details of the derivation. 

For the Lorenz gauge $\partial_\mu \mathscr{A}^\mu=0$, the GF+FP term is given by
\begin{align}
  \mathscr{L}_{\text{GF+FP}} 
 =  % - \partial_\mu \mathscr{B} \cdot \mathscr{A}^\mu 
  \mathscr{B} \cdot \partial_\mu \mathscr{A}^\mu 
  + \frac{1}{2} \alpha \mathscr{B} \cdot \mathscr{B} - i \partial^\mu \bar{\mathscr{C}} \cdot \mathscr{D}_\mu \mathscr{C} %\nonumber\\
 = - i {\boldsymbol \delta}_{\rm{B}}  \left[ \bar{\mathscr{C}} \cdot  \left( \partial^\mu \mathscr{A}_\mu + \frac{\alpha}{2} \mathscr{B} \right) \right] ,
  \label{eq:Lorenz-GF+FP}
\end{align}
where $\alpha$ is the gauge-fixing parameter.
The change of the Lagrangian density under the generalized local gauge transformation is given by $\alpha$-independent expression:
\begin{align}
 		\delta^\omega \mathscr{L}_{\rm GF+FP}(x)
%=  i  {\boldsymbol \delta}_{\rm{B}}  \bar{{\boldsymbol \delta}}_{\rm{B}}  \mathscr{A}_\mu(x) \cdot  \partial^\mu \omega (x)
		 = i  \bm{\delta}_{\rm B} (\mathscr{D}_\mu \bar{\mathscr{C}}(x))^A  \partial^\mu \omega^A(x) .
\end{align}
In the Lorenz gauge, the above WT relation (\ref{CCF-WTI_gen1}) reduces to 
\begin{align}
  &\int d^D x e^{i p (x - y)} \partial^x_\mu \braket{ {\rm T}^{} {\mathscr{J}_\omega^{\mu }}_\nu^A (x) \partial^\nu \omega^A(x) \mathscr{A}_\lambda^B (y)}  
  \nonumber\\
  = & i \partial_\lambda \omega^B (y) + \int d^D x e^{i p (x - y)} \partial^\mu \omega^A(x) \braket{ {\rm T}^{} i  \bm{\delta}_B (\mathscr{D}_\mu \bar{\mathscr{C}}(x))^A \mathscr{A}_\lambda^B(y)} .
  \label{CCF-WTI_gen_Lorenz}
\end{align}
The second term of \eqref{CCF-WTI_gen_Lorenz}  is rewritten using 
$\bm{\delta}_{\rm B}( \mathscr{D}_\mu \bar{\mathscr{C}})=\bm{\delta}_{\rm B}( \partial_\mu \bar{\mathscr{C}} + g(\mathscr{A}_\mu \times \bar{\mathscr{C}}) )
=i \partial_\mu \mathscr{B}+  g\bm{\delta}_{\rm B}(\mathscr{A}_\mu \times \bar{\mathscr{C}})  $ as
\begin{align}
  &\int d^D x e^{i p (x - y)} \partial^\mu \omega^A(x) \braket{ {\rm T}^{} i    \bm{\delta}_{\rm B} (\mathscr{D}_\mu \bar{\mathscr{C}}(x))^A \mathscr{A}_\lambda^B (y)} \nonumber\\
  = &-\int d^D x e^{i p (x - y)} \partial^\mu \omega^A(x) \partial_\mu^x \braket{ {\rm T}^{} \mathscr{B}^A (x) \mathscr{A}_\lambda^B (y)} \nonumber\\
  &+ i \int d^D x e^{i p (x - y)} \partial^\mu \omega^A (x) \braket{ {\rm T}^{} g \bm{\delta}_{\rm B} (\mathscr{A}_\mu \times \bar{\mathscr{C}})^A (x) \mathscr{A}_\lambda^B (y)} \nonumber\\
  = &-  \int d^D x e^{i p (x - y)} \partial^\mu \omega^A(x) \partial_\mu^x \braket{ {\rm T}^{} \mathscr{B}^A (x) \mathscr{A}_\lambda^B (y)} \nonumber\\
  &+ i \int d^D x e^{i p (x - y)} \partial^\mu \omega^A (x) \braket{ {\rm T}^{} g (\mathscr{A}_\mu \times \bar{\mathscr{C}})^A (x) \bm{\delta}_{\rm B} \mathscr{A}_\lambda^B (y)} \nonumber\\
  = &-   \int d^D x e^{i p (x - y)} \partial^\mu \omega^A(x) \partial_\mu^x \braket{ {\rm T}^{} \mathscr{B}^A (x) \mathscr{A}_\lambda^B (y)} \nonumber\\
  &+ i \int d^D x e^{i p (x - y)} \partial^\mu \omega^A (x) \braket{ {\rm T}^{} g (\mathscr{A}_\mu \times \bar{\mathscr{C}})^A (x) (\mathscr{D}_\lambda \mathscr{C})^B (y)}
  \nonumber\\
  = &-  \int d^D x e^{i p (x - y)} \partial^\mu \omega^A(x) \partial_\mu^x  i \frac{\partial_{\lambda}^x }{\partial_x^2} \delta^D(x-y) \delta^{AB}  \nonumber\\
  &+ i \int d^D x e^{i p (x - y)} \partial^\mu \omega^A (x)
  \left(g_{\mu\lambda}- \frac{\partial_{\mu}^x \partial_{\lambda}^x}{\partial_x^2} \right) u^{AB}(x-y) ,
  \label{eq:WI_Lorenz_general_2nd}
\end{align}
where we have used $\braket{\bm{\delta}_{\rm B} F} = 0$ for any functional $F$ due to the physical state condition,  
the exact form of the propagator in the Lorenz gauge
%
\begin{comment}
\footnote{
The relation (\ref{CCF-DCC00}) is derived as follows. 
Remember the Slavnov-Taylor identity holds, 
\begin{align}
  \langle 0 | {\rm T} ( \mathscr{D}_{\mu} \mathscr{C} )^A(x) i \bar{\mathscr{C}}^B(y) | 0 \rangle 
= \langle 0 | {\rm T} \mathscr{A}_{\mu}^A(x) \mathscr{N}^B(y) | 0 \rangle ,
\end{align}
which follows from 
\begin{align}
  \langle 0 | \{ iQ_B, {\rm T} \mathscr{A}_{\mu}^A(x) i \bar{\mathscr{C}}^B(y) \} | 0 \rangle 
= 0.
\end{align}
}
\end{comment}
%
\begin{align}
 \langle 0 | {\rm T} \mathscr{A}_{\mu}^A(x) \mathscr{B}^B(y) | 0 \rangle=
\langle 0 | {\rm T} (\mathscr{D}_{\mu} \mathscr{C})^A(x) i \bar{\mathscr{C}}^B(y) | 0 \rangle
=&  i \frac{\partial_{\mu}^x }{\partial_x^2} \delta^D(x-y) \delta^{AB} ,
\label{CCF-DCC00}
\end{align}
and 
the definition of the \textbf{Kugo-Ojima (KO) function}\index{Kugo-Ojima (KO) function} $u^{AB}$ in the configuration space
\begin{align}
\langle 0 | {\rm T} (\mathscr{D}_{\mu} \mathscr{C})^A(x) (g \mathscr{A}_{\nu} \times \bar{\mathscr{C}})^B(y) | 0 \rangle
= \left(g_{\mu\nu}- \frac{\partial_{\mu}^x \partial_{\nu}^x}{\partial_x^2} \right) u^{AB}(x-y) .
\label{CCF-KO-func-real}
\end{align}

Thus, we arrive at the desired general condition in the Lorenz gauge written in the Euclidean form:
\begin{align}
  \lim_{p \rightarrow \ 0}  \int d^D x e^{i p (x - y)} \partial_\mu \omega^A (x) \left (\delta_{\mu \lambda} -  \frac{\partial_{\mu}^x \partial_{\lambda}^x}{\partial_x^2} \right) %\nonumber\\  &
    \left[ \delta^D(x-y)\delta^{AB} + u^{AB}(x-y) \right]  = 0 .
  \label{eq:condition_Lorenz_general}
\end{align}
%where we have used the color symmetry to write $u^{AB}(x-y)=\delta^{AB}u(x-y)$.

This confinement criterion can be applied to the Abelian and non-Abelian gauge theory as well irrespective of the compact or non-compact formulation, 
and is able to understand confinement in all the cases. 

In the non-compact gauge theory formulated in terms of the Lie-algebra-valued gauge field, the choice of $\omega^A (x)$ as the non-compact variable linear in $x$,
\begin{align}
\omega^A (x) = \text{const.} + \epsilon_\mu^A x_\mu =\text{const. + non-compact~variable} ,
\end{align}
is allowed. 
Indeed, for this choice, the criterion (\ref{eq:condition_Lorenz_general}) is reduced to
\begin{align}
  \epsilon_\mu^A \lim_{p \rightarrow \ 0}  \left( \delta_{\mu \lambda} - \frac{p_\mu p_\lambda}{p^2} \right) 
 \left[ \delta^{AB}  + \tilde u^{AB} (p) \right] = 0.
  \label{eq:condition_Lorenz_special}
\end{align}
This reproduces the KO condition $\tilde u^{AB}(0) = -\delta^{AB}$ as first shown by Hata.% \cite{Hata82}. 

In the compact gauge theory formulated in terms of the group-valued gauge field, on the other hand, we must choose the compact, namely, angle variables for $\omega^A$, 
\begin{align}
\omega^A (x) =\text{const. + angle~variable}=\text{const. + compact~variable} .
\end{align}
In this case, the KO condition is not needed to be satisfied.  Thus, the KO condition is neither sufficient nor necessary for confinement in the compact gauge theory.

To see this difference clearly,  consider the Abelian gauge theory.
For the Abelian gauge theory, the KO function is identically zero $u^{AB}(x-y) \equiv 0$, i.e., $u(0)=0$. Therefore, the KO condition is not satisfied, which means that confinement does not occur. 
Indeed, this result is true and  only true in the non-compact Abelian gauge theory. 
In the compact gauge theory, however, confinement does occur even in the Abelian gauge theory, as is well known in the lattice gauge theory \cite{Creutz83}. 
This case is also understood by using the above criterion. 
In this case, we must choose the compact $\omega$. Then the criterion (\ref{eq:condition_Lorenz_general}) is indeed shown to be satisfied and confinement does occurs, although $u(0)=0$, as given in the next section. 

 In the MA gauge, we must choose compact $\omega^j$. 
 If we choose the non-compact $\omega^j$, then the KO condition is not satisfied for the diagonal components in agreement with the previous works \cite{SS83,HN93}. However, this does not mean that confinement does not occur.  This merely means that the choice of $\omega$ is wrong to discuss confinement in the MA gauge.  In the MA gauge, the criterion (\ref{eq:condition_Lorenz_general}) is indeed satisfied and confinement occurs, as shown in the next section.

%\newpage
%%%%%%%%%%%%%%%%%%%%%%%%%%%%%%%%%%%%%%%%%%%%%%%%%%%%%%%%%
%%%%%%%%%%%%%%%%%%%%%%%%%%%%%%%%%%%%%%%%%%%%%%%%%%%%%%%%%
\section{Residual gauge symmetry  in the MA gauge}
%%%%%%%%%%%%%%%%%%%%%%%%%%%%%%%%%%%%%%%%%%%%%%%%%%%%%%%%%
%%%%%%%%%%%%%%%%%%%%%%%%%%%%%%%%%%%%%%%%%%%%%%%%%%%%%%%%%

%%%%%%%%%%%%%%%%%%%%%%%%%%%%%%%%%%%%%%%%%%%%%%%%%%%%%%%%%
\subsection{The general case of the residual gauge symmetry in the MA gauge}
%%%%%%%%%%%%%%%%%%%%%%%%%%%%%%%%%%%%%%%%%%%%%%%%%%%%%%%%%

We try to find a color confinement condition in the MA gauge by performing the similar analyses to the Lorenz gauge. 
We decompose the Lie-algebra valued quantity to the diagonal Cartan part and the remaining off-diagonal part, e.g., the gauge field $\mathscr{A}_\mu=\mathscr{A}_\mu^A T_A$ with the generators $T_A$ ($A=1,\ldots,N^2-1$) of the Lie algebra $su(N)$ has the decomposition:
\begin{align}
  \mathscr{A}_\mu(x)=\mathscr{A}_\mu^A(x) T_A
  = a_\mu^j(x) H_j + A_\mu^a(x) T_a ,
  \label{eq:decomp}
\end{align}
where $H_j$ are the Cartan generators and $T_a$ are the remaining generators of the Lie algebra $su(N)$. 
In what follows, the indices $j,k,\ell,\ldots$ label the diagonal components and the indices $a,b,c,\ldots$ label  the off-diagonal components. 
The maximal Abelian (MA) gauge is given by
\begin{align}
  (\mathscr{D}_{}^\mu [a] A_\mu(x))^a := \partial^\mu A_\mu^a(x) + {} gf^{ajb}a^\mu{}^j(x) A_\mu^b(x)
  = 0 ,
  \label{eq:MA-condition}
\end{align}
where we have used $f^{j k A} = 0$.
%Here we have introduced a parameter which interporate the MA gauge ${}=1$ and the Lorenz gauge ${}=0$. 
The MA gauge is a partial gauge which fix the off-diagonal components, but does not fix the diagonal components.  Therefore, we further  impose the Lorenz gauge for the diagonal components
\begin{align}
  \partial^\mu a_\mu^j(x) = 0 .
  \label{eq:MA-Lorenz}
\end{align}
The GF+FP term for the gauge-fixing condition \eqref{eq:MA-condition} and \eqref{eq:MA-Lorenz} with respective gauge-fixing parameter $\alpha$ and $\beta$ is given 
using the BRST transformation as 
\begin{align}
  \mathscr{L}_{\text{GF+FP}} 
  = &- i \bm{\delta}_{\rm B}\left\{ \bar{C}^a \left(\mathscr{D}_{}^\mu [a] A_\mu + \frac{\alpha}{2} B\right)^a \right\} 
  - i \bm{\delta}_{\rm B} \left\{ \bar{c}^j \left(\partial^\mu a_\mu + \frac{\beta}{2} b\right)^j \right\} ,
  \label{eq:MA-GF+FP1}
\end{align}
which reads
\begin{align}
  \mathscr{L}_{\text{GF+FP}} = &- (\mathscr{D}_{}^\mu [a]^{b a} B^a) A_\mu^b + \frac{\alpha}{2} B^a B^a - i (\mathscr{D}_{}^\mu [a]^{b a} \bar{C}^a) \mathscr{D}_\mu [a]^{b c} C^c \nonumber\\
  &- i g (\mathscr{D}_{}^\mu [a]^{b a} \bar{C}^a) f^{b c d} A_\mu^c C^d - i g (\mathscr{D}_{}^\mu [a]^{b a} \bar{C}^a) f^{b c j} A_\mu^c c^j \nonumber\\
  &+ i {} g \bar{C}^a f^{a j b} \partial_\mu c^j A^{\mu b} + i {} g^2 \bar{C}^a f^{a j b} f^{j c d} A_\mu^c C^d A^{\mu b}  \nonumber\\
  &- \partial^\mu b^j a_\mu^j+ \frac{\beta}{2} b^j b^j - i \partial^\mu \bar{c}^j \partial_\mu c^j - i g \partial^\mu \bar{c}^j f^{j a b} A_\mu^a C^b .
  \label{eq:MA-GF+FP2}
\end{align}

The local gauge transformation of the Lagrangian has the following form as shown in Appendix A 
%\begin{align}
%  \delta^\omega \mathscr{L}_{\text{GF + FP}} = &i \bm{\delta}_B \partial_\mu \bar{c}^j \partial^\mu \omega^j + i \bm{\delta}_B (\mathscr{D}_\mu [\mathscr{A}] \bar{\mathscr{C}})^a \partial^\mu \omega^a + i \bm{\delta}_B (\partial^\mu \mathscr{D}_\mu [\mathscr{A}] \bar{\mathscr{C}})^a \omega^a ,
%\end{align}
\begin{align}
		\delta^\omega \mathscr{L} &= \delta^\omega \mathscr{L}_{\text{GF+FP}}  
		= \partial_\mu \mathscr{J}^\mu_\omega 
%= \partial_\mu \mathscr{J}_\omega^\mu{}_\nu^A \partial^\nu \omega^A
%\nonumber\\  
 = 
 g \partial_{\mu} \mathscr{J}^{\mu}  \cdot \omega 
+ \left[    \partial_{\nu} \mathscr{F}^{\mu \nu}   + g \mathscr{J}^{\mu}  \right] \cdot \partial_{\mu} \omega 
\nonumber\\  
&= 
 g \partial^\mu {J}_{\mu}^j \omega^j 
+ \left[ \partial^{\nu} {f}_{\mu \nu}^j + g {J}_{\mu}^j \right]  \partial_{\mu} \omega^j
+ g \partial^\mu {J}_{\mu}^a \omega^a  
+ \left[ \partial^{\nu} {F}_{\mu \nu}^a + g {J}_{\mu}^a \right]  \partial_{\mu} \omega^a
\nonumber\\  
%- \mathscr{F}^{\mu \nu}(x) \cdot \partial_{\mu} \partial_{\nu} \omega(x) 
&= 
%-  \partial^\mu \partial_\mu b^j \omega^j  
%- \partial_\mu b^j \partial^\mu \omega^j  
i \bm{\delta}_B \partial_\mu \bar{c}^j \partial^\mu \omega^j
+ i {\boldsymbol \delta}_{\rm{B}} \partial^\mu (\mathscr{D}_{\mu}[\mathscr{A}]\bar{\mathscr{C}})^a \omega^a  
+ i {\boldsymbol \delta}_{\rm{B}} (\mathscr{D}_{\mu}[\mathscr{A}]\bar{\mathscr{C}})^a \partial^\mu \omega^a  
.
		\label{eq:MA-current-divergence2}
\end{align}
The diagonal global current is conserved in the MA gauge, in contrast to the off-diagonal current which is not conserved.
This is BRST exact,  which shows that the local gauge current $\mathscr{J}^\mu_\omega$  is conserved in the physical state space. 

The WT relation in the MA gauge can be calculated in the similar way to the Lorenz gauge by using \eqref{eq:MA-current-divergence2} as follows.
%\eqref{eq:MA-current-divergence-diagonal} and \eqref{eq:MA-current-divergence-offdiagonal}
\begin{align}
 & \int d^D x \ e^{i p (x - y)} \partial_\mu^x \braket{ {\rm T}^{} \mathscr{J}_\omega^\mu (x) \mathscr{A}_\lambda^B (y)} \nonumber\\
  = & i \int d^D x \ e^{i p (x - y)} \delta^D (x - y)  \braket{\delta^\omega \mathscr{A}_\lambda^B(y)} 
+ \int d^D x \  e^{i p (x - y)} \braket{ {\rm T}^{} \partial_\mu \mathscr{J}_\omega^\mu (x) \mathscr{A}_\lambda^B (y)}   \nonumber\\
%= & i \braket{(\mathscr{D}_\lambda [\mathscr{A}]\omega)^B(y)} 
%+ \int d^D x \  e^{i p (x - y)} \braket{ {\rm T}^{} \partial^x_\mu {\mathscr{J}_\omega^{\mu}}_\nu^A (x) \partial^\nu \omega^A(x) \mathscr{A}_\lambda^B (y)} \nonumber\\
  = & i \braket{(\mathscr{D}_\lambda [\mathscr{A}]\omega)^B(y)}  
%+  \int d^D x \ e^{i p (x - y)}  \omega^j(x) \braket{ {\rm T}^{} i \bm{\delta}_{\rm B} \{ \partial^\nu  \partial_\nu \bar{c}^j (x)\} \mathscr{A}_\lambda^B (y)} 
%\nonumber\\&
+  \int d^D x \ e^{i p (x - y)}  \partial^\nu \omega^j(x)  \braket{ {\rm T}^{} i \bm{\delta}_{\rm B} \{ \partial_\nu \bar{c}^j (x)\} \mathscr{A}_\lambda^B (y)} \nonumber\\
%&+  \int d^D x \ e^{i p (x - y)} \omega^a(x) \braket{ {\rm T}^{} i \bm{\delta}_{\rm B} \{ \partial^\nu (\mathscr{D}_\nu [\mathscr{A}] \bar{\mathscr{C}})^a(x)  \} \mathscr{A}_\lambda^B (y)} \nonumber\\
 &+  \int d^D x \ e^{i p (x - y)} [\partial^\nu \omega^a(x) + \omega^a(x) \partial^\nu _{x}] \braket{ {\rm T}^{} i \bm{\delta}_{\rm B} \{ (\mathscr{D}_\nu [\mathscr{A}] \bar{\mathscr{C}})^a(x)  \} \mathscr{A}_\lambda^B (y)} \nonumber\\
= &  i \partial_\lambda \omega^B (y) 
%-  \int d^D x \ e^{i p (x - y)}  \omega^j(x) \partial^\nu \braket{ {\rm T}^{}   \partial_\nu b^j (x) \mathscr{A}_\lambda^B (y)} \nonumber\\  &
  -  \int d^D x \ e^{i p (x - y)} \partial^\nu \omega^j(x)  \braket{ {\rm T}^{}   \partial_\nu b^j (x) \mathscr{A}_\lambda^B (y)} \nonumber\\
%&+  \int d^D x \ e^{i p (x - y)}  \omega^a(x) \braket{ {\rm T}^{}   (\partial^\nu \mathscr{D}_\nu [\mathscr{A}] \bar{\mathscr{C}})^a(x)   i\bm{\delta}_{\rm B} \mathscr{A}_\lambda^B (y)}   \nonumber\\
  &+  \int d^D x \ e^{i p (x - y)} [\partial^\nu \omega^a(x) + \omega^a(x) \partial^\nu _{x}] \braket{ {\rm T}^{}   (\mathscr{D}_\nu [\mathscr{A}] \bar{\mathscr{C}})^a(x)   i\bm{\delta}_{\rm B} \mathscr{A}_\lambda^B (y)} 
  \nonumber\\
= &  i \partial_\lambda \omega^B (y) 
  -  \int d^D x \ e^{i p (x - y)} \partial^\nu \omega^j(x)  \partial_\nu^x \braket{ {\rm T}^{}    b^j (x) \mathscr{A}_\lambda^B (y)} \nonumber\\
  &+  \int d^D x \ e^{i p (x - y)} [\partial^\nu \omega^a(x) + \omega^a(x) \partial^\nu _{x}]  \braket{ {\rm T}^{}   (\mathscr{D}_\nu [\mathscr{A}] \bar{\mathscr{C}})^a(x)   i(\mathscr{D}_\lambda [\mathscr{A}] \mathscr{C})^B (y)} 
    \nonumber\\
= &  i \partial_\lambda \omega^B (y) 
  -  \int d^D x \ e^{i p (x - y)} \partial^\nu \omega^j(x)  \partial_\nu^x \braket{ {\rm T}^{}    b^j (x) \mathscr{A}_\lambda^B (y)} \nonumber\\
  &+  \int d^D x \ e^{i p (x - y)} [\partial^\nu \omega^a(x) + \omega^a(x) \partial^\nu _{x}]  \partial_\nu^x \braket{ {\rm T}^{}   \bar{{C}}^a(x) i(\mathscr{D}_\lambda [\mathscr{A}] \mathscr{C})^B (y)} 
 \nonumber\\
  &+  \int d^D x \ e^{i p (x - y)} [\partial^\nu \omega^a(x) + \omega^a(x) \partial^\nu _{x}] \braket{ {\rm T}^{}   g( \mathscr{A}_\nu \times \bar{\mathscr{C}})^a(x)   i(\mathscr{D}_\lambda [\mathscr{A}] \mathscr{C})^B (y)}  .
  \label{eq:WI-MA}
\end{align}

We focus on the diagonal gauge field $a_\lambda^k$. 
For $B=k$, the second term of \eqref{eq:WI-MA}  reads
\begin{align}
   %\int d^D x e^{i p (x - y)} \braket{ {\rm T}^{} \{Q_B , \partial_\nu \bar{c}^j (x)\} a_\lambda^k (y)} %\nonumber\\
  \int d^D x \ e^{i p (x - y)} \partial^\nu \omega^j(x)  \braket{ {\rm T}^{} \partial_\nu b^j (x) a_\lambda^k (y)} %\nonumber\\   &
  =   \int d^D x \  e^{i p (x - y)} \partial^\nu \omega^j(x)  \partial_\nu^x i\frac{\partial_{\lambda}^x }{\partial_x^2} \delta^D(x-y) \delta^{jk} ,%\nonumber\\
%&=  i  \frac{p_\nu p_\lambda}{p^2} \partial^\nu \omega^k(y)  ,
   %=  -   i p_\nu \frac{p_\lambda}{p^2} \delta^{j k} ,
   \label{WT-MAG-diag}
\end{align}
where we have used 
the exact form of the propagator in the MA gauge
\begin{align}
 \langle 0 | {\rm T}^{}  b^j (x) a_\mu^k (y) | 0 \rangle 
=   i \frac{\partial_{\mu}^x }{\partial_x^2} \delta^D(x-y) \delta^{jk} ,
\label{CCF-DCC-MAG}
\end{align}
which is shown to hold in the MA gauge as  in the Lorenz gauge. 
This follows from the relation which is proved in the Appendix B:
\begin{align}
\langle 0 | {\rm T} i \mathscr{B}^A(x) \mathscr{A}_{\mu}^k(y) | 0 \rangle 
= \langle 0 | {\rm T} \bar{\mathscr{C}}^A(x)  ( \mathscr{D}_{\mu} \mathscr{C} )^k(y)   | 0 \rangle 
%= - \frac{\partial_\mu^y}{\partial_y^2}  \delta^D(x-y) \delta^{Ak}  
= \frac{\partial_\mu^x}{\partial_x^2}  \delta^D(x-y) \delta^{Ak}  .
\end{align}
This means that the third term of \eqref{eq:WI-MA} vanishes, $\langle 0 | {\rm T} \bar{{C}}^a(x)  ( \mathscr{D}_{\mu} \mathscr{C} )^k(y)   | 0 \rangle=0$.
Moreover, we show the following transversality in the Appendix B
\begin{align}
\langle 0 | {\rm T} (g \mathscr{A}_{\nu} \times \bar{\mathscr{C}})^A(x) (\mathscr{D}_{\mu}[\mathscr{A}] \mathscr{C})^k(y)  | 0 \rangle
= \left(g_{\mu\nu}- \frac{\partial_{\mu}^x \partial_{\nu}^x}{\partial_x^2} \right) v^{Ak}(x-y) ,
\label{CCF-tra-MAG}
\end{align}
where we defined the KO function $v^{Ak}(x-y)$ in the MA gauge. 
If the color symmetry is unbroken, we have $v^{Ak}(x-y)=\delta^{Ak} v(x-y)$, which means that $v^{ak}(x-y)=0$. Therefore, the last term  of \eqref{eq:WI-MA} vanishes. 
If the color symmetry is broken and is not recovered, (i) we expect that the 2-point function between the diagonal component and the off-diagonal component  rapidly decreases to zero in the long-distance limit $|x-y| \to \infty$, corresponding to $p \to 0$, or  
(ii) we set $\omega^a(x)=0$ and $\omega^j(x) \not=0$ from the beginning, since we do not need to treat the diagonal and off-diagonal components on equal footing in this case. 
%Therefore, we find that the diagonal components for $B=k$ obey

Consequently, we obtain the condition  for the restoration of the residual local gauge symmetry for the diagonal gauge field
\begin{align}
& \lim_{p \rightarrow \ 0} \int d^D x \ e^{i p (x - y)} \partial_\mu^x \braket{ {\rm T}^{} \mathscr{J}_\omega^\mu (x) {a}_\lambda^k (y)}  \nonumber\\
&=
  \lim_{p \rightarrow \ 0}  i \int d^D x \ e^{i p (x - y)}  \partial^\mu \omega^k(x) 
% ( \delta_{\mu\lambda} \Box_D-\partial_\mu \partial_\lambda) \Box_D^{- 1} (x , y) 
   (\delta_{\nu \lambda} - \partial_\nu \partial_\lambda  \Box_D^{- 1}) %\nonumber\\  &
    \delta^D(x-y)   = 0 ,
  \label{eq:condition_MAG_diag_general}
\end{align}
where $\Box_D^{- 1} (x , y)$ denotes the Green function of the  Laplacian $\Box_D=\partial_\mu \partial_\mu$ in the $D$-dimensional Euclidean space. 
By using integration by part, the above condition is rewritten into 
\begin{align}
& \lim_{p \rightarrow \ 0} \int d^D x \ e^{i p (x - y)} \partial_\mu^x \braket{ {\rm T}^{} \mathscr{J}_\omega^\mu (x) {a}_\lambda^k (y)}  \nonumber\\
&=
  \lim_{p \rightarrow \ 0}  p^\mu \int d^D x \ e^{i p (x - y)}    \omega^k(x) 
  ( \delta_{\mu\lambda} \Box_D-\partial_\mu \partial_\lambda) \Box_D^{- 1} (x , y) 
       = 0 .
  \label{eq:condition_MAG_diag_general2}
\end{align}
However, the previous expression (\ref{eq:condition_MAG_diag_general}) is more useful for checking the condition because $\partial^\mu \omega^k(x)$ has the universal form for any $D$, while this is not the case for $\omega^k(x)$ itself. 
\subsection{Restoration of the residual gauge symmetry due to singular topological configurations}
%%%%%%%%%%%%%%%%%%%%%%%%%%%%%%%%%%%%%%%%%%%%%%%%%%%%%%%%%

\begin{comment}
In the above, we have estimated  the WT relation for the choice $\omega = \epsilon_\mu x^\mu$ as the residual gauge symmetry according to the previous study \cite{Hata82}.
%in the manifestly Lorentz covariant operator formalism of the non-Abelian gauge theory due to Kugo and Ojima \cite{manifestly-Lorenz-covariant, color-confinement-condition}. 
In the MA gauge, however, the imposition of disappearance of the massless pole in the WT relation lead to different conditions depending on the component: a non-trivial condition (\ref{criterion-offdiag}) for the off-diagonal components and a trivial one (\ref{criterion-diag}) for the diagonal components. 
Consequently, the local gauge symmetry can be restored for the off-diagonal components, while the restoration does not occur for the diagonal components. 
%This speciality of the MA gauge agrees with the observation  \cite{renormalization-group-flow} given based on the renormalization group.
Does this mean that the color confinement criterion of the KO type cannot be applied to the MA gauge as claimed in \cite{SS83,HN93}? 
\end{comment}

From now on, we reconsider  the issue of restoration of the residual local gauge symmetry in the MA gauge. 
First, we recall the fact that the local gauge current is conserved in the physical subspace irrespective of the choice of $\omega$. 
Therefore, there are no special reasons to choose the specific form $\omega = \epsilon_\mu x^\mu$ linear  in the coordinate $x^\mu$ as the residual gauge symmetry, in sharp contrast to the Abelian case where it is the exact residual symmetry in the whole state space. 
Second, it has been shown that the configurations responsible for quark confinement in the MA gauge are singular configurations with non-trivial topology, which is represented by the infrared Abelian dominance and magnetic monopole dominance. 
%The investigation so far for the symmetry restoration were done for the residual gauge symmetry linear in $x$. 
In view of these, we do not restrict the residual local gauge symmetry to the linear type and allow singular choice to examine whether the residual local gauge symmetry associated with certain singular topological configurations can be restored to make the true vacuum disordered, even if it is broken in the perturbative vacuum.
If such configurations for $\omega$ exist, we can guess that they contribute to confinement, according to the experience in the Lorenz gauge. 
The restoration is probed by the disappearance of the massless pole in the WT relation.

If we choose $\omega^j(x) = \epsilon_\nu^j x^\nu$, this indeed reproduces non-vanishing divergent result.  % (\ref{criterion-diag}).
However, this choice must be excluded  in the MA gauge, since the maximal torus subgroup $U(1)^{N-1}$  for the diagonal components of the gauge field is a compact subgroup of the original $SU(N)$ group which is indeed a compact non-Abelian gauge group. 
However, the choice $\omega^j(x) = \epsilon_\nu^j x^\nu$ does not reflect this fact. 
In some sense, $\omega^j(x)$ must be angle variables reflecting the compactness of the gauge group.

For concreteness, we consider the $SU(2)$ case with singular configurations coming from the angle variables. 
In what follows, we work in the Euclidean space and use subscripts instead of the Lorentz indices. 
Then, $|x-y|^2$ denotes the squared Euclidean distance $|x-y|^2:=\sum_{\mu=1}^{D} |x_\mu-y_\mu|^2$.
As the residual gauge transformation, we take the following examples which satisfy both the Lorenz gauge condition $\partial_\mu \mathscr{A}_\mu^A=0$ and the MA gauge condition $(\mathscr{D}_\mu [a] A_\mu)^a =0$ (and $\partial^\mu a_\mu^j=0$).  

For $D=2$, a collection of vortices of Abrikosov-Nielsen-Olesen type \cite{NO79}
%\footnote{
%H. B. Nielsen   and P. Olesen,
%A Quantum Liquid Model for the QCD Vacuum: Gauge and Rotational Invariance of Domained and Quantized Homogeneous Color Fields,
%Nucl. Phys. B{\bf 160},  380--396 (1979) 
%}
\begin{align}
   \partial_\mu \omega^j (x)
   = \sum_{s=1}^{n} C_s \varepsilon_{j\mu\nu} \frac{(x-a_s)_\nu}{|x-a_s|^2} \ (j=3, \ \mu,\nu=1,2)  \ (x,a_s \in \mathbb{R}^2) ,
   \label{sing-2}
\end{align}
where $C_s$  ($s=1,\ldots,n$) are  arbitrary constants.
This type of $\omega(x)$ is indeed an angle variable $\theta$ going around a point $a=(a_1, a_2) \in \mathbb{R}^2$, because 
\begin{align}
   \omega (x) = \theta(x) 
   =: \arctan \frac{x_2-a_2}{x_1-a_1} 
   \Longrightarrow 
   \partial_\mu \omega (x) 
   = - \varepsilon_{\mu\nu} \frac{x_\nu-a_\nu}{(x_1-a_1)^2+(x_2-a_2)^2} \ (\mu=1,2).
   \label{sing-2b}
\end{align}
This is a topological configuration which is classified by the winding number of the map from the circle in the space to the circle in the target space: $S^1 \to U(1) \cong S^1$, i.e., by the first Homotopy group $\pi_1(S^1)=\mathbb{Z}$. 

For $D=3$, a collection of magnetic monopoles of the Wu-Yang type \cite{WY75}, 
%\footnote{
%Tai Tsun Wu, Chen Ning Yang, 
%Concept of Nonintegrable Phase Factors and Global Formulation of Gauge Fields,
%Phys. Rev. D{\bf 12}, 3845%--3857
% (1975). 
%
%T.T. Wu and C.N. Yang,
%Dirac Monopole Without Strings: Monopole Harmonics, 
%Nucl. Phys. B{\bf 107}, 365--380  (1976).
%T.T. Wu and C.N. Yang,
%Dirac's Monopole Without Strings: Classical Lagrangian Theor,  
%Tai Tsun Wu, Chen Ning Yang,
%Phys. Rev. D{\bf 14}, 437%--445
%  (1976).
%}
which corresponds to the zero size limit of the `t Hooft-Polyakov magnetic monopole \cite{tHP74}
%\footnote{
%G. 't Hooft,
%Magnetic Monopoles in Unified Gauge Theories,
%Nucl. Phys. B{\bf 79}, 276%--284
% (1974).
%
%A. M. Polyakov,
%Particle Spectrum in the Quantum Field Theory,
%JETP Lett. {\bf 20}, 194%--195
% (1974).
%Pisma Zh. Eksp. Teor. Fiz. {\bf 20}, 430--433  (1974). 
%}
\begin{align}
   \partial_\mu \omega^j (x)
   = \sum_{s=1}^{n}  C_s  \varepsilon_{j\mu\nu} \frac{(x-a_s)_\nu}{|x-a_s|^2} \ (j=3, \ \mu,\nu=1,2,3)  \ (x,a_s \in \mathbb{R}^3) .
   \label{sing-3}
\end{align}
The Wu-Yang magnetic monopole is gauge equivalent to the Dirac magnetic monopole, which means that they are mutually related by the gauge transformation, see Appendix C. 
A magnetic monopole is a topological configuration which is classified by the winding number of the map from the sphere in the space to the sphere in the target space: $S^2 \to SU(2)/U(1) \cong S^2$, i.e., by the second Homotopy group $\pi_2(S^2)=\mathbb{Z}$. 

For $D=4$,  a collection of merons of Alfaro-Fubini-Furlan \cite{AFF76,CDG78}, 
%\footnote{
%V. De Alfaro, S. Fubini and G. Furlan,
%A new classical solution of the Yang-Mills field equations,
%Phys. Lett. B{\bf 65}, 163%--166
% (1976).
% \\
%}
instantons of the Belavin-Polyakov-Shwarts-Tyupkin (BPST) type \cite{BPST75} in the non-singular gauge with zero size, 
%\footnote{
%A.A. Belavin, A. M. Polyakov, A.S. Shwarts and  Yu.S. Tyupkin,
%Pseudoparticle Solutions of the Yang-Mills Equations,
%Phys. Lett. B{\bf 59}, 85--87 (1975). 
%}
\begin{align}
   \partial_\mu \omega^j (x)
   = \sum_{s=1}^{n}  C_s \eta^{j}_{\mu\nu} \frac{(x-a_s)_\nu}{|x-a_s|^2} \ (j=3, \ \mu,\nu=1,2,3,4)  \ (x,a_s \in \mathbb{R}^4) .
   \label{sing-4}
\end{align}
For the mutual relationship among instanton solutions from our point of view, see Appendix C. 
Meron and instanton are  topological configuration which are classified by the winding number of the map from the 3-dimensional sphere in the space to the sphere in the target space: $S^3 \to SU(2) \cong S^3$, i.e., by the third Homotopy group $\pi_3(S^3)=\mathbb{Z}$. 
%\footnote{
For the details on the topological soliton solutions, see e.g., \cite{Rajaraman}.
%R. Rajaraman,
%Solitons and Instantons: An Introduction to Solitons and %Instantons in Quantum Field Theory
%(North-Holland, Amsterdam, 1987).
%
%N. Manton and P. Sutcliffe,
%Topological Solitons
%(Cambridge Univ. Press, 2007).
%}
It has been shown that the meron and instanton configurations can be the source for generating   magnetic monopole loops for $D=4$ Euclidean space \cite{KFSS08}.

By taking into account the antisymmetry $\varepsilon^{j}_{\mu\nu}=-\varepsilon^{j}_{\nu\mu}$, $\eta^{j}_{\mu\nu}=-\eta^{j}_{\nu\mu}$, it is easy to show that all these configurations satisfy the Laplace equation $\Box \omega^j(x)=0$ almost everywhere except for the locations  $a_s \in \mathbb{R}^D$ ($s=1,\ldots,n$) of the singularities: $\Box \omega^j(x)= \sum_{s=1}^{n} C_s  \delta^D(x-a_s)$. 
These configurations are examples of the classical solutions of the Yang-Mills field equation with non-trivial topology.
%\footnote{
For a review on the classical solutions of the Yang-Mills field equation, see e.g., \cite{Actor79}.
%}
In particular, the above configuration for $D=4$ is suggested from the solution of the Yang-Mills field equation obtained by adopting the Corrigan-Fairlie-'tHooft-Wilczek (CFtHW) Ansatz \cite{CF77,tHooft76,Wilczek77}:
\begin{align}
%  \mathscr{A}_\mu(x) =   \frac{\sigma_A}{2}  
\mathscr{A}_\mu^A(x) 
=    \eta^A_{\mu\nu} \partial_\nu \ln \phi(x)  %f_\nu(x) , \quad f_\nu(x) := \partial_\nu \ln \phi(x) 
 ,
\end{align}
where $\eta^A_{\mu\nu}$ %=\eta^{(+)}{}^A_{\mu\nu}$
 is the symbol defined by 
\begin{align}
 \eta^A_{\mu\nu} %\equiv \eta^{(+)}{}^A_{\mu\nu} 
 := \epsilon_{A\mu\nu 4} + \delta_{A\mu}\delta_{\nu 4} - \delta_{\mu 4}\delta_{A \nu}
 = \begin{cases}
    \epsilon_{Ajk} & (\mu=j, \nu=k) \cr
    \delta_{Aj} & (\mu=j, \nu= 4) \cr
    -\delta_{Ak} & (\mu=4, \nu=k)
   \end{cases}
 .
\end{align}
\begin{comment}
Similarly, we can define
$\bar{\eta}^A_{\mu\nu}=:\eta^{(-)}{}^A_{\mu\nu}$  as
\begin{align}
 \bar{\eta}^A_{\mu\nu} \equiv  \eta^{(-)}{}^A_{\mu\nu} := \epsilon_{A\mu\nu 4} - \delta_{A\mu}\delta_{\nu 4} + \delta_{\mu 4}\delta_{A \nu}
 = \begin{cases}
    \epsilon_{Ajk} & (\mu=j, \nu=k) \cr
    -\delta_{Aj} & (\mu=j, \nu= 4) \cr
    +\delta_{Ak} & (\mu=4, \nu=k)
   \end{cases}
 .
\end{align}
\end{comment}
The Yang-Mills field of this Ansatz becomes the solution \cite{Rajaraman} when $\phi(x)$ satisfies 
\begin{align}
   \Box \phi(x)
= 0 .
\end{align}
Note that $\eta^A_{\mu\nu}$ is self-dual, i.e., $\eta^A_{\mu\nu}=*\eta^A_{\mu\nu}:=\frac12 \epsilon_{\mu\nu\alpha\beta} \eta^A_{\alpha\beta}$.
%, while $\bar{\eta}^A_{\mu\nu}$ is anti-selfdual, i.e., $-\bar{\eta}^A_{\mu\nu}=*\bar{\eta}^A_{\mu\nu}$.
%
The Yang-Mills field in the  CFtHW  Ansatz satisfies the Lorenz gauge condition:
\begin{align}
 \partial_\mu \mathscr{A}_\mu^A(x) 
 =  \eta^A_{\mu\nu} \partial_\mu f_\nu(x) 
 =  \eta^A_{\mu\nu} \partial_\mu \partial_\nu \ln \phi(x)
= 0 
 ,
\end{align}
where we have used the antisymmetry $\eta^A_{\mu\nu}=-\eta^A_{\nu\mu}$.
Simultaneously, it satisfies the MA gauge condition: 
\begin{align}
\mathscr{D}_\mu[\mathscr{A}^3] \mathscr{A}_\mu^{\pm}(x):= (\partial_\mu - ig \mathscr{A}_\mu^3) (\mathscr{A}_\mu^1(x) \pm i \mathscr{A}_\mu^2(x)) = 0
,
\end{align}
where we have used
$
\eta^A_{\mu\alpha} \eta^B_{\mu\beta}
= \delta_{AB}\delta_{\alpha\beta}+\epsilon_{ABC}\eta^C_{\alpha\beta}
$.

%\begin{ex}
%Verify that the configurations given in (\ref{sing-2}), (\ref{sing-3}), and (\ref{sing-4}) satisfy the Laplace equation $\Box \omega^j(x)=0$ almost everywhere except for the locations  $a_s \in \mathbb{R}^D$ ($s=1,\ldots,n$) of the singularities. 
%\end{ex}

%\begin{ex}
%Verify that the configurations $\omega(x)$ given in (\ref{sing-3}), and (\ref{sing-4}) are indeed angle variables in the sense similar to (\ref{sing-2}).  
%\end{ex}

We can show that the restoration condition is satisfied for these singular configurations:
\begin{align}
  \lim_{p \to 0}   \int d^D x \ e^{i p (x - y)} \frac{(x-a_s)_\nu}{|x-a_s|^2}  \left( \delta_{\mu\lambda} \Box_D -\partial_\mu \partial_\lambda  \right)   \frac{\frac{\Gamma \left(\frac{D}{2} - 1\right)}{4 \pi^{D / 2}}}{(|x - y|^2)^{\frac{D - 2}{2}}}
  =0  .
  \label{eq:MA-WI-singular2}
\end{align}
where we have used the expression of  the Green function $\Box_D^{- 1} (x , y)$  of the  Laplacian $\Box_D=\partial_\mu \partial_\mu$ in the $D$-dimensional Euclidean space  given by
\begin{align}
  \Box^{- 1}_D (x , y) = \int \frac{d^D p}{(2 \pi)^D} e^{i p (x - y)} 
\frac{1}{- p^2} = - \frac{\Gamma \left(\frac{D}{2} - 1\right)}{4 \pi^
{D / 2}} \frac{1}{|x - y|^{D - 2}} ,
  \label{Green-Laplacian}
\end{align}
where $\Gamma$ is the gamma function with the integral representation given by 
\begin{align}
  \Gamma (z) = \int_0^\infty dt \ t^{z - 1} e^{- t} \ (z > 0) .
\end{align}
For any $D \ge 2$, this integral (\ref{eq:MA-WI-singular2}) goes to zero linearly in $p$ in the limit $p \to 0$.
Therefore, the restoration of the residual local gauge symmetry occurs. 
The details on the calculation of the integral are given in the Appendix C. 

From the above considerations, we find immediately that the compact U(1) gauge theory can have the disordered confinement phase in which the residual local gauge symmetry is restored, while the non-compact U(1) gauge theory has the deconfined Coulomb phase in which the residual local gauge symmetry remains broken.

%%%%%%%%%%%%%%%%%%%%%%%%%%%%%%%%%%%%%%%%%%%%%%%%%%%%%%%%%%%%%%%%%%%%%%
%\section{Reconsideration of the Abelian case}
%%%%%%%%%%%%%%%%%%%%%%%%%%%%%%%%%%%%%%%%%%%%%%%%%%%%%%%%%%%%%%%%%%%%%%

\section{Conclusion and discussion}

In this paper, we have reexamined the restoration of the residual local gauge symmetry left even after imposing the gauge fixing condition in quantum gauge field theories. 
According to  \cite{Hata82}, the restoration condition of the residual local gauge symmetry in the Lorenz gauge is equivalent to the KO color confinement condition ensuring that all colored particles including dynamical quarks and gluons are confined.  However, it has been reported \cite{SS83,HN93} that this correspondence is unfortunately invalid in the MA gauge, although the dual superconductor picture based on condensation of magnetic monopoles works well to verify quark confinement as confirmed by many works studied in the MA gauge. 
In order to resolve this controversial issue and make a bridge between the KO color confinement and the dual superconductor picture for quark confinement, we have investigated a generalization of the color confinement criterion.

An important lesson we obtained in this work to understand color confinement in quantum gauge theories is that the compactness and non-compactness must be discriminated for the local gauge transformation of the gauge field. 
We have found that the KO color confinement  criterion can be applied only to the non-compact gauge theory. 
In the non-compact formulation of the quantum gauge theory, indeed, the Kugo-Ojima criterion is expected to give a reasonable condition to be satisfied for realizing color confinement. 
However, this is not the case for the compact formulation of the quantum gauge theory in which the residual local gauge transformation must be defined for compact angle variables which correspond to the non-trivial topological configurations of the gauge field expressed by the singular gauge field configurations.
This is the reason why the KO criterion obtained first in the Lorenz gauge cannot be applied to the MA gauge, since the maximal torus group is a compact group.

From this viewpoint, we have shown that the restoration of the residual local gauge symmetry  indeed occurs in the MA gauge for the SU(N) Yang-Mills theory in two-, three- and four-dimensional Euclidean spacetime once the singular topological configurations of gauge fields, i.e., vortices, magnetic monopoles, merons/instantons are respectively taken into account. 
As a byproduct, we notice immediately that the compact U(1) gauge theory can have the disordered confinement phase, while the non-compact U(1) gauge theory has the deconfined Coulomb phase. 
Our results suggest that the color confining phase in the compact gauge theory is a disordered phase caused by non-trivial topological configurations irrespective of the gauge choice. 
This understanding of confinement agrees with the results obtained by the lattice gauge theory based the compact formulation.

In our treatment of the residual local gauge symmetry, we assumed the existence of the BRST symmetry.  However, it is well known that the usual gauge fixing condition is not sufficient to completely fix the gauge, since there are so many Gribov copies satisfying the given gauge fixing condition. This is indeed the case for both in the Lorenz and the MA gauges. 
Therefore, it is necessary to examine to what extent the results obtained in this paper are affected by the existence of the Gribov copies \cite{Kondo09}.  

In this paper we have discussed confined phase and deconfined (Coulomb) phase. When the matter field is included, there could exist the Higgs phase in which the original massless gauge fields become massive due to the Brout-Englert-Higgs (BEH) mechanism by absorbing the massless Nambu-Goldstone bosons which appear associated with the spontaneous breaking of the gauge  symmetry.  We wish to say something about the gauge-independent BEH mechanism \cite{Kondo16} from the viewpoint of the restoration of the residual local gauge symmetry. 
In distinguishing the Higgs phase from the confinement phase, it is important to exhaust all possible cases  \cite{GM17}: the global gauge symmetry is broken or unbroken, at the same time the residual local gauge symmetry is restored or remains broken. 

Moreover, it is important to extend our investigations to finite temperature case to see how there exists the expected phase transition separating the deconfinement phase at high temperature from the confinement phase at low temperature. 
These important issues are to be tackled in near future.

\section*{Acknowledgment}

This work was  supported by Grant-in-Aid for Scientific Research, JSPS KAKENHI Grant Number (C) No.19K03840.

% can use a bibliography generated by BibTeX as a .bbl file
% BibTeX documentation can be easily obtained at:
% http://www.ctan.org/tex-archive/biblio/bibtex/contrib/doc/

%\bibliographystyle{ptephy}
%\bibliography{sample}
%
% once the .bbl file has been generated then place the text in your article.

\vspace{0.2cm}
\noindent

%This is added by T. Yoneya (editor-in-chief) on 2020/07/09.

\let\doi\relax

%without this code before the command "\begin{thebibliography}{}" , an error will be %flagged. When the bibliography is provided as separate .bib file, then this code %should be placed above the commands "\bibliographystyle{}" and "\bibliography{}" %inside the main TeX file. 

\begin{thebibliography}{9}
\bibitem{dualsuper}
%\bibitem{Nambu74}
Y. Nambu,
%Strings, monopoles, and gauge fields,
Phys. Rev. D{\bf 10}, 4262%--4268
 (1974).
\\
G. 't Hooft,
  in: High Energy Physics, edited by A. Zichichi 
(Editorice Compositori, Bologna, 1975).
\\
%\bibitem{Mandelstam76}
S. Mandelstam,
%Vortices and quark confinement in non-abelian gauge theories, 
Phys. Report  {\bf 23}, 245%--249
 (1976).


\bibitem{CP97}
M.N. Chernodub, M.I. Polikarpov,
Abelian projections and monopoles, 
Lectures given at NATO Advanced Study Institute on Confinement, Duality and Nonperturbative Aspects of QCD, Cambridge, England, 23 Jun - 4 Jul 1997,  
[arXiv:hep-th/9710205] 


\bibitem{KKSS15}
K.-I. Kondo, S. Kato, A. Shibata and T. Shinohara,
%Quark confinement: dual superconductor picture based on a non-Abelian Stokes theorem and reformulations of  Yang-Mills theory, 
Phys. Rept. \textbf{579}, 1%--226
 (2015) 
[arXiv:1409.1599 [hep-th]].


\bibitem{HK21}
Y. Hayashi and K.-I. Kondo,
%Reconstructing propagators of confined particles in the presence of complex singularities,
Phys. Rev.D \textbf{104}, 074024 (2021).
arXiv: 2105.07487 [hep-th]


%\bibitem{Greensite03}
% J. Greensite,
%The Confinement problem in lattice gauge theory.
%Prog. Part. Nucl. Phys. {\bf 51}, 1 (2003)  
%[arXiv:hep-lat/0301023]]   
 
 
\bibitem{KO79}
 T. Kugo and I. Ojima,
%Local covariant operator formalism of non-Abelian gauge theories and quark confinement problem,
Suppl. Prog. Theor. Phys. \textbf{66}, 1--130  (1979).
\\
T. Kugo and I. Ojima, 
Phys. Lett. \textbf{73B}, 459 (1978); 
Prog. Theor. Phys. \textbf{60}, 1869 (1978); 61, 294, 644 (1979).


\bibitem{Kugo89}
T. Kugo, 
Quantum theory of gauge fields, I, II, 
%vol.~I: 1--272,  vol.~II: 1--284, 
in Japanese
(Baifu-kan, Tokyo, 1989).


%\bibitem{NO90}
%N. Nakanishi and I. Ojima,  
%Covariant operator formalism of gauge theories and quantum gravity
%(World Scientific, Singapore, 1990).
%World Sci. Lect. Notes Phys. 27, 1--434 (1990). 


\bibitem{Hata82}
H. Hata, 
Prog. Theor. Phys. \textbf{67}, 1607 (1982).


\bibitem{Hata83}
H. Hata, 
%Restoration of the Local Gauge Symmetry and Color Confinement in Non-Abelian Gauge Theories. II, 
Prog. Theor. Phys. \textbf{69}, 1524% E1536
 (1983).


\bibitem{Kugo95}
T. Kugo, 
%The Universal renormalization factors Z(1)/Z(3) and color confinement condition in nonAbelian gauge theory,
arXiv: hep-th/9511033


\bibitem{Wilson74}
G. Wilson,
%Confinement of Quarks
Phys. Rev. D{\bf 10}, 2445 %--2459
 (1974).
%CLNS-262
%DOI: 10.1103/PhysRevD.10.2445


\bibitem{MAG}
%\bibitem{KLSW87}
A. Kronfeld, M. Laursen, G. Schierholz and U.-J. Wiese,
%Monopole condensation and color confinement,
Phys. Lett. B{\bf 198}, 516%--520
 (1987).  


\bibitem{KondoII}
K.-I. Kondo, 
%Yang-Mills theory as a deformation of topological field theory, dimensional reduction and quark confinement,
%  Chiba Univ. Preprint, CHIBA-EP-103,
Phys. Rev. D {\bf 58}, 105019 (1998).
arXiv: [hep-th/9801024] 


\bibitem{Kondo01}
K.-I. Kondo,
%Vacuum condensate of mass dimension 2 as the origin of mass gap and quark confinement,
Phys. Lett. B{\bf 514}, 335--345 (2001). 
[hep-th/0105299] 
\\
K.-I. Kondo,
%A Physical meaning of mixed gluon ghost condensate of mass dimension two,
Phys. Lett. B{\bf 572}, 210%--215
 (2003). 
arXiv: [hep-th/0306195]  


\bibitem{SS83}
T. Suzuki and K. Shimada, 
%CONFINEMENT CRITERIA AND COMPACT QED in (2+1)-dimensions,
Prog. Theor. Phys.  \textbf{69}, 1537%--1547
 (1983). 


\bibitem{Polyakov77}
A.M. Polyakov,
%Quark confinement and topology of gauge theories,
Nucl. Phys. B{\bf 120}, 429%--458
 (1977).


\bibitem{HN93}
H. Hata and I. Niigata, 
Nucl. Phys. \textbf{B389}, 440 (1993).


\bibitem{Polyakov75}
A.M. Polyakov,
%Compact gauge fields and the infrared catastrophe,
Phys. Lett. B{\bf 59}, 82%--84
 (1975).


\bibitem{Elitzur75}
S. Elitzur, 
%Impossibility of Spontaneously Breaking Local Symmetries, 
Phys. Rev. D\textbf{12}, 3978%--3982  
 (1975).  
\\
G. F. De Angelis, D. de Falco, and F. Guerra, 
%A Note on the Abelian Higgs-Kibble Model on a Lattice: Absence of Spontaneous Magnetization, 
Phys.Rev. D\textbf{17}, 1624%--1628  
 (1978).  


\bibitem{FP71}
R. Ferrari and L. E. Picasso, 
Nucl. Phys. \textbf{B31}, 316 (1971).

\bibitem{KTU85}
T. Kugo, H. Terao and S. Uehara,
%DYNAMICAL GAUGE BOSONS AND HIDDEN LOCAL SYMMETRIES, 
Prog. Theor. Phys. Suppl. 85, 122 %--135
 (1985) .
%Contribution to: Meson 50: Kyoto International Symposium: The Jubilee of the Meson Theory, 122-135

\bibitem{Creutz83}
M. Creutz, 
Quarks, gluons and lattice 
(Cambridge Univ. Press, 1983).

%\bibitem{MM94}
%I. Montvay and G. M\"unster,
%Quantum Fields on a Lattice
%(Cambridge Univ. Press, 1994).


\bibitem{NO79}
 H. B. Nielsen   and P. Olesen,
%A Quantum Liquid Model for the QCD Vacuum: Gauge and Rotational Invariance of Domained and Quantized Homogeneous Color Fields,
Nucl. Phys. B{\bf 160},  380%--396
 (1979) 


\bibitem{WY75}
Tai Tsun Wu, Chen Ning Yang, 
%Concept of Nonintegrable Phase Factors and Global Formulation of Gauge Fields,
Phys. Rev. D{\bf 12}, 3845%--3857
 (1975). 
%
T.T. Wu and C.N. Yang,
%Dirac Monopole Without Strings: Monopole Harmonics, 
Nucl. Phys. B{\bf 107}, 365--380  (1976).
T.T. Wu and C.N. Yang,
%Dirac's Monopole Without Strings: Classical Lagrangian Theor,  
%Tai Tsun Wu, Chen Ning Yang,
Phys. Rev. D{\bf 14}, 437%--445
  (1976).


\bibitem{tHP74}
G. 't Hooft,
%Magnetic Monopoles in Unified Gauge Theories,
Nucl. Phys. B{\bf 79}, 276%--284
 (1974).
%
A. M. Polyakov,
%Particle Spectrum in the Quantum Field Theory,
JETP Lett. {\bf 20}, 194%--195
 (1974).
%Pisma Zh. Eksp. Teor. Fiz. {\bf 20}, 430--433  (1974). 


\bibitem{AFF76}
V. De Alfaro, S. Fubini and G. Furlan,
%A new classical solution of the Yang-Mills field equations,
Phys. Lett. B{\bf 65}, 163%--166
 (1976).


%\bibitem{AFF77}
%V. De Alfaro, S. Fubini and G. Furlan,
%Properties of O(4)xO(2) symmetric solutions of the Yang-Mills field equations,
%Phys. Lett. B{\bf 72}, 203%--207
% (1977).

\bibitem{BPST75}
A.A. Belavin, A. M. Polyakov, A.S. Shwarts and  Yu.S. Tyupkin,
%Pseudoparticle Solutions of the Yang-Mills Equations,
Phys. Lett. B{\bf 59}, 85%--87
 (1975). 


\bibitem{Rajaraman}
R. Rajaraman,
Solitons and Instantons: An Introduction to Solitons and Instantons in Quantum Field Theory
(North-Holland, Amsterdam, 1987).
\\
N. Manton and P. Sutcliffe,
Topological Solitons
(Cambridge Univ. Press, 2007).
\\
Y.M. Shnir, 
Magnetic monopoles 
(Springer, 2005).


\bibitem{CF77}
E. Corrigan and D.B. Fairlie,
Phys. Lett. B {\bf 67}, 69 (1977).


\bibitem{tHooft76}
G. 't Hooft,
%Computation of the Quantum Effects Due to a Four-Dimensional Pseudoparticle,
Phys.Rev. D{\bf 14}, 3432-3450 (1976); Erratum-ibid. D{\bf 18}, 2199 (1978). 


\bibitem{Wilczek77}
F. Wilczek,
in ``Quark Confinement and Field Theory'' eds. by D. Stump and D. Weingarten, (Wiley, New York, 1977).


\bibitem{Actor79}
A. Actor,
%Classical solutions of SU(2) Yang-Mills theories,
Rev. Mod. Phys. {\bf 51}, 461%--525
 (1979).


\bibitem{CDG78}
C.G. Callan, R. Dashen and D.J. Gross,
%Toward a Theory of the Strong Interactions,
Phys. Rev. D{\bf 17}, 2717%--2763
 (1978).


\bibitem{KFSS08}
K.-I. Kondo, N. Fukui, A. Shibata and T. Shinohara, 
%Magnetic Monopole Loops supported by a Meron pair as the Quark Confiner,
%CHIBA-EP-172, KEK-2008-14,
Phys. Rev. D{\bf 78}, 065033 (2008). 
arXiv:0806.3913 [hep-th]  


%\bibitem{FKSS10}
N. Fukui, K.-I. Kondo, A. Shibata, and T. Shinohara, 
%Jackiw-Nohl-Rebbi two-instanton as a source of magnetic monopole loop, 
%CHIBA-EP-183-KEK-PREPRINT-2010-8 
Phys. Rev. D{\bf 82},  045015 (2010). 
arXiv:1005.3157 [hep-th] 


%\bibitem{FKSS12}
N. Fukui, K.-I. Kondo, A. Shibata, and T. Shinohara, 
%Magnetic monopole loops generated from two-instanton solutions: Jackiw-Nohl-Rebbi versus 't Hooft instanton, 
%CHIBA-EP-193-KEK-PREPRINT-2012-9 
Phys. Rev. D{\bf 86},  065020 (2012). 
arXiv:1205.4972 [hep-th]  


\bibitem{Kondo09}
K.-I. Kondo,
%Kugo-Ojima color confinement criterion and Gribov-Zwanziger horizon condition,
Phys. Lett. B\textbf{678}, 322%--330
 (2009). 
arXiv:0904.4897 [hep-th]  


\bibitem{Kondo16}
K.-I. Kondo,
%Gauge-invariant description of Higgs phenomenon and quark confinement, 
Phys. Lett. B\textbf{762}, 219-224 (2016). 
arXiv:1606.06194 [hep-th]


\bibitem{GM17}
J. Greensite and K. Matsuyama,
%Confinement criterion for gauge theories with matter fields, 
Phys. Rev. D\textbf{96}, 094510 (2017).
e-Print: 1708.08979 [hep-lat]

\end{thebibliography}

\appendix
\section{Generalized local transformation}

First, we define the (infinitesimal) local transformation $\delta^H$ with $x$-dependent parameters 
$\omega(x)=\omega^A(x)T_A$:
\begin{align}
		\delta^H \Phi(x) =
\begin{cases}
		g\Phi(x) \times \omega(x), \quad &{\rm for} \quad \Phi = \mathscr{A}_{\mu}, \mathscr{C}, \bar{\mathscr{C}} , \mathscr{N} , \\
		ig \omega(x) \varphi(x), \quad &{\rm for\ the\ matter\ field\ } \Phi = (\varphi)_a .
\end{cases}
\end{align}
For the gauge field $\mathscr{A}_{\mu}$ and the matter field $\Phi$ (except for the ghost field $\mathscr{C}$, antighost field $\bar{\mathscr{C}}$ and the Nakanishi-Lautrup field $\mathscr{N}$), this transformation $\delta^H$ becomes identical with the \textbf{global gauge transformation} or \textbf{color rotation} when the parameters $\omega(x)$ are set to be $x$-independent.

In addition, we define the \textbf{generalized local gauge transformation} $\delta^\omega$ by
\begin{align}
		\delta^\omega \Phi(x) =
\begin{cases}
		\mathscr{D}_{\mu}[\mathscr{A}] \omega(x) = \partial_{\mu} \omega(x) +   \delta^H \mathscr{A}_{\mu}(x),
		\quad {\rm for} \quad \Phi = \mathscr{A}_{\mu} , \\
		  \delta^H \Phi(x), \quad {\rm for} \quad \Phi = \mathscr{C}, \bar{\mathscr{C}}, \mathscr{N} , \varphi,
\end{cases}
\end{align}
where $\delta^\omega$ differs from $\delta^H$ only by the inhomogeneous term $\partial_{\mu} \omega(x)$ added to $g \delta^H \mathscr{A}_{\mu}(x)$ for the gauge field $\mathscr{A}_{\mu}$.
For the gauge field $\mathscr{A}_{\mu}$ and the matter field $\Phi$, the transformation $\delta^\omega$ is nothing but the local gauge transformation.

Next, we consider the change of the total Lagrangian $\mathscr{L}=\mathscr{L}(\Phi, \partial\Phi)$ under the generalized local gauge transformation $\delta^\omega$:
\begin{align}
%		\delta^\omega \mathscr{L} 
  \delta^\omega \mathscr{L}(\Phi, \partial\Phi) 
%\nonumber\\
		 &=  \mathscr{L}(\Phi + \delta^\omega \Phi , \partial\Phi + \partial\delta^\omega \Phi) - \mathscr{L}(\Phi, \partial\Phi)
%		&= \frac{\partial \mathscr{L}}{\partial \Phi} \cdot \delta^\omega \Phi
%		+ \frac{\partial \mathscr{L}}{\partial (\partial_{\mu} \Phi)} \cdot \delta^\omega (\partial_{\mu} \Phi) 
\nonumber\\
		&= \frac{\partial \mathscr{L}}{\partial \Phi} \cdot \delta^\omega \Phi
		+ \frac{\partial \mathscr{L}}{\partial \partial_{\mu} \Phi} \cdot \partial_{\mu} \delta^\omega \Phi \nonumber\\
		&= \delta^H \mathscr{L} + \frac{\partial \mathscr{L}}{\partial \mathscr{A}_{\mu}} \cdot \partial_{\mu} \omega
		+ \frac{\partial \mathscr{L}}{\partial (\partial_{\mu} \mathscr{A}_{\nu})} \cdot \partial_{\mu} \partial_{\nu} \omega ,
\end{align}
where
\begin{align}
		\delta^H \mathscr{L} := \frac{\partial \mathscr{L}}{\partial \Phi} \cdot   \delta^H \Phi
		+ \frac{\partial \mathscr{L}}{\partial (\partial_{\mu} \Phi)} \cdot   \partial_{\mu} \delta^H  \Phi .
\end{align}
By separating the parameter $\omega$ from $\delta^H \Phi$,
\begin{align}
		\delta^H \Phi^A(x) = gG^{AB}[\Phi] \omega^B(x) ,
\end{align}
which yields
\begin{align}
%\delta^H \partial_{\mu} \Phi^A(x) = 
\partial_{\mu} \delta^H \Phi^A(x)
		= g\partial_{\mu} G^{AB}[\Phi]  \omega^B(x) + gG^{AB}[\Phi]  \partial_{\mu} \omega^B(x) ,
\end{align}
$\delta^H \mathscr{L}$ is written as
\begin{align}
		\delta^H \mathscr{L}
		= \left[ \frac{\partial \mathscr{L}}{\partial \Phi^A} g G^{AB}
		+ \frac{\partial \mathscr{L}}{\partial (\partial_{\mu} \Phi^A)} g \partial_{\mu} G^{AB} \right] \omega^B
		+ \frac{\partial \mathscr{L}}{\partial (\partial_{\mu} \Phi^A)} g G^{AB} \partial_{\mu} \omega^B .
		\label{CCF-dHL}
\end{align}
The second term on the right-hand side of (\ref{CCF-dHL}) includes a part which is nothing but the Noether current associated
with the global gauge transformation with $x$-independent $\omega(x) \equiv \omega$:
\begin{align}
		\mathscr{J}^{\mu B}(x) = \frac{\partial \mathscr{L}}{\partial (\partial_{\mu} \Phi^A(x))} G^{AB} [\Phi(x)]  .
\end{align}
Then the  first term on the right-hand side of (\ref{CCF-dHL}) is rewritten as 
\begin{align}
		\frac{\partial \mathscr{L}}{\partial \Phi} \cdot gG + \frac{\partial \mathscr{L}}{\partial (\partial_{\mu} \Phi)} \cdot g \partial_{\mu} G
		= \frac{\delta S}{\delta \Phi}   \cdot  gG + g \partial_{\mu} \mathscr{J}^{\mu} ,
\end{align}
where we have defined
\begin{align}
% 		[\mathscr{L}] := 
\frac{\delta S}{\delta \Phi(x)} 
:= \frac{\partial \mathscr{L}}{\partial \Phi(x)} - \partial_{\mu} \left(\frac{\partial \mathscr{L}}{\partial (\partial_{\mu} \Phi(x))} \right) .
\end{align}
Thus (\ref{CCF-dHL}) is rewritten into
\begin{align}
		\delta^H \mathscr{L}(x) = \left[ g \partial_{\mu} \mathscr{J}^{\mu}(x) 
	+ \frac{\delta S}{\delta \Phi(x)}  \cdot gG  \right] 	\cdot \omega(x) + g \mathscr{J}^{\mu}(x) \cdot \partial_{\mu} \omega(x) ,
\end{align}
and hence $\delta^\omega \mathscr{L}$ reads
\begin{align}
		\delta^\omega \mathscr{L}(x) =& \left[ g \partial_{\mu} \mathscr{J}^{\mu}(x) 
	+ \frac{\delta S}{\delta \Phi(x)}  \cdot gG  \right] 	\cdot \omega(x)
 \nonumber\\
		&+ \left[ \frac{\partial \mathscr{L}}{\partial \mathscr{A}_{\mu}(x)}  + g \mathscr{J}^{\mu}(x)  \right] \cdot \partial_{\mu} \omega(x)
		+ \frac{\partial \mathscr{L}}{\partial (\partial_{\mu} \mathscr{A}_{\nu}(x))} \cdot \partial_{\mu} \partial_{\nu} \omega(x) .
\label{CCF-dGL}
\end{align}
	By introducing the functional derivative of the action $S$ with respect to the gauge field 
\begin{align}
   \frac{\delta S}{\delta \mathscr{A}_{\mu}(x)} 
:= \frac{\partial \mathscr{L}}{\partial \mathscr{A}_{\mu}(x)} - \partial_{\nu} \left(\frac{\partial \mathscr{L}}{\partial (\partial_{\nu} \mathscr{A}_{\mu}(x))} \right)  ,
\end{align}
the relation (\ref{CCF-dGL}) is rewritten as
\begin{align}
		\delta^\omega \mathscr{L}(x)  =& 
		\left[ g \partial_{\mu} \mathscr{J}^{\mu}(x) 
	+ \frac{\delta S}{\delta \Phi(x)}  \cdot gG  \right] 	\cdot \omega(x)
		\nonumber\\
		&+ \left[  \frac{\delta S}{\delta \mathscr{A}_{\mu}(x)}  + \partial_{\nu} \frac{\partial \mathscr{L} }{\partial (\partial_{\nu} \mathscr{A}_{\mu}(x))} + g \mathscr{J}^{\mu}(x) \right] \cdot \partial_{\mu} \omega(x) \nonumber\\
		& + \frac{\partial \mathscr{L} }{\partial (\partial_{\mu} \mathscr{A}_{\nu}(x))} \cdot \partial_{\mu} \partial_{\nu} \omega(x) .
		\label{SSB-id}
\end{align}
By taking into account the Yang-Mills action, we have
\begin{align}
\frac{\partial \mathscr{L} }{\partial (\partial_{\nu} \mathscr{A}_{\mu})}
=\frac{\partial \mathscr{L}_{\rm YM} }{\partial (\partial_{\nu} \mathscr{A}_{\mu})}
= \mathscr{F}^{\mu \nu} = - \mathscr{F}^{\nu \mu}  ,
\end{align}
which yields 
\begin{align}
		\delta^\omega \mathscr{L}(x)  =& 
		\left[ g \partial_{\mu} \mathscr{J}^{\mu}(x) 
	+ \frac{\delta S}{\delta \Phi(x)}  \cdot gG  \right] 	\cdot \omega(x)
		\nonumber\\
		&+ \left[  \frac{\delta S}{\delta \mathscr{A}_{\mu}(x)}  + \partial_{\nu} \mathscr{F}^{\mu \nu}(x)  + g \mathscr{J}^{\mu}(x) \right] \cdot \partial_{\mu} \omega(x) \nonumber\\
		& - \mathscr{F}^{\mu \nu}(x) \cdot \partial_{\mu} \partial_{\nu} \omega(x) .
		\label{SSB-id2}
\end{align}
It should be remarked that this relation has been obtained without using the equation of motion and other on mass-shell relations. 

The on mass-shell relation is obtained by using the equation of motion: 
\begin{align}
		 \frac{\delta S}{\delta \Phi(x)}  :=
		\frac{\partial \mathscr{L}}{\partial \Phi(x)} - \partial_{\mu} \left(\frac{\partial \mathscr{L}}{\partial (\partial_{\mu} \Phi(x))} \right) = 0,
\end{align}	
which includes especially the equation of motion for the Yang-Mills field
\begin{align}
\frac{\delta S}{\delta \mathscr{A}_{\mu}^A(x)}  = \frac{\partial \mathscr{L} }{\partial \mathscr{A}_{\mu}^A(x)}
		- \partial_{\nu} \left(\frac{\partial \mathscr{L} }{\partial (\partial_{\nu} \mathscr{A}_{\mu}^A(x))} \right)  = 0 .
\end{align}
On mass-shell, thus,  taking into account equations of motion for the fields,
we have
\begin{align}
		\delta^\omega \mathscr{L}(x)  = 
		 g \partial_{\mu} \mathscr{J}^{\mu}(x) \cdot \omega(x)
%\nonumber\\		&
+ \left[    \partial_{\nu} \mathscr{F}^{\mu \nu}(x)  + g \mathscr{J}^{\mu}(x) \right] \cdot \partial_{\mu} \omega(x)
%\nonumber\\ & 
%- \mathscr{F}^{\mu \nu}(x) \cdot \partial_{\mu} \partial_{\nu} \omega(x) .
		\label{SSB-id3}
\end{align}

We apply the above relation for the generalized local gauge transformation to the total Lagrangian   $\mathscr{L} =\mathscr{L}^{\rm tot}_{\rm YM}$:
\begin{align}
%\mathscr{L}^{\rm tot}_{\rm YM} 
		\mathscr{L} = \mathscr{L}_{\rm inv} + \mathscr{L}_{\rm GF+FP} .%+ \mathscr{L}_m .
\end{align}
%The ordinary Yang-Mills theory is reproduced in the limit $M \rightarrow 0$. 
The original Lagrangian $\mathscr{L}_{\rm inv}$ is invariant by construction under the generalized local gauge transformation:
\begin{align}
 \delta^\omega \mathscr{L}_{\rm inv} = 0.
\end{align}
This means that the generalized local gauge transformation of the Lagrangian is determined only by that of the GF+FP term alone:
\begin{align}
 \delta^\omega \mathscr{L}_{\rm inv} = 0
 \Rightarrow \delta^\omega \mathscr{L}=\delta^\omega \mathscr{L}_{\rm GF+FP}  .
\end{align}
In order to discuss the generalized local gauge transformation of the GF+FP term, we observe the commutativity between the  gauge transformation ${\boldsymbol \delta}^G$ and the BRST-transformation ${\boldsymbol \delta}_{\rm{B}}$ (anti-BRST transformation $\bar{{\boldsymbol \delta}}_{\rm{B}}$):
\begin{align}
 [{\boldsymbol \delta}^G, {\boldsymbol \delta}_{\rm{B}}] = 0 = [{\boldsymbol \delta}^G, \bar{{\boldsymbol \delta}}_{\rm{B}}] ,
		\label{SSB-comm}
\end{align}
which can be checked to hold. 
%In particular, we find 
%\begin{align}
%	[{\boldsymbol \delta}^G, {\boldsymbol \delta} ] = 0 = [{\boldsymbol \delta}^G, \bar{{\boldsymbol \delta}} ]  .
%\label{SSB-comm2}
%\end{align}

\subsection{Lorenz gauge}

For a simple choice of the GF+FP term for the Lorenz gauge:
\begin{align}
		\mathscr{L}_{\rm GF+FP}
		&= - i {\boldsymbol \delta}_{\rm{B}}  \left[ \bar{\mathscr{C}} \cdot  \left( \partial^\mu \mathscr{A}_\mu + \frac{\alpha}{2} \mathscr{N} \right) \right]
 ,
\end{align}
its change under the generalized local gauge transformation is calculated as
\begin{align}
 		\delta^\omega \mathscr{L}_{\rm GF+FP}
 		=&     - i \delta^\omega {\boldsymbol \delta}_{\rm{B}}  \left[ \bar{\mathscr{C}} \cdot  \left( \partial^\mu \mathscr{A}_\mu + \frac{\alpha}{2} \mathscr{N} \right) \right]\nonumber\\
		=&     - i  {\boldsymbol \delta}_{\rm{B}}  \delta^\omega \left[ \bar{\mathscr{C}} \cdot   \partial^\mu \mathscr{A}_\mu + \frac{\alpha}{2} \bar{\mathscr{C}} \cdot   \mathscr{N}   \right]\nonumber\\
		=&     - i  {\boldsymbol \delta}_{\rm{B}}   \left[ \delta^\omega( \bar{\mathscr{C}} \cdot   \partial^\mu \mathscr{A}_\mu ) + \frac{\alpha}{2} \delta^\omega( \bar{\mathscr{C}} \cdot   \mathscr{N}  )  \right]\nonumber\\
		=& - i  {\boldsymbol \delta}_{\rm{B}}   \left[ - \partial^\mu \delta^\omega  \bar{\mathscr{C}}  \cdot    \mathscr{A}_\mu   -  \partial^\mu  \bar{\mathscr{C}} \cdot   \delta^\omega  \mathscr{A}_\mu  \right] 
 \nonumber\\
		=& - i  {\boldsymbol \delta}_{\rm{B}}   \left[ - \partial^\mu (g\bar{\mathscr{C}}  \times \omega) \cdot    \mathscr{A}_\mu   -  \partial^\mu  \bar{\mathscr{C}} \cdot   (\partial_\mu \omega + g \mathscr{A}_\mu \times \omega) \right]
 \nonumber\\
		=&  - i  {\boldsymbol \delta}_{\rm{B}}    [ 
-  ( \partial^\mu\bar{\mathscr{C}}  \times \omega) \cdot g\mathscr{A}_\mu   
-  ( \bar{\mathscr{C}}  \times \partial^\mu\omega) \cdot  g\mathscr{A}_\mu   
 \nonumber\\
		&-   \partial^\mu  \bar{\mathscr{C}} \cdot  \partial_\mu \omega - \partial^\mu  \bar{\mathscr{C}} \cdot (g \mathscr{A}_\mu \times \omega) 
 ]
 \nonumber\\
		=&  - i  {\boldsymbol \delta}_{\rm{B}}   \left[ 
 -  ( \bar{\mathscr{C}}  \times \partial_\mu\omega) \cdot  g\mathscr{A}^\mu   
-   \partial^\mu  \bar{\mathscr{C}} \cdot  \partial_\mu \omega  
\right]
 \nonumber\\
		=&   i  {\boldsymbol \delta}_{\rm{B}}   \left( 
     \partial^\mu  \bar{\mathscr{C}}  + g\mathscr{A}^\mu \times \bar{\mathscr{C}}        
 \right)  \cdot  \partial_\mu \omega
\nonumber\\
		 =&  i  {\boldsymbol \delta}_{\rm{B}}   \left( 
     \mathscr{D}^\mu[\mathscr{A}]  \bar{\mathscr{C}}   
 \right)  \cdot  \partial_\mu \omega
\nonumber\\
		 =&  i  {\boldsymbol \delta}_{\rm{B}}  \bar{{\boldsymbol \delta}}  \mathscr{A}^\mu \cdot  \partial_\mu \omega
  .
	\label{SSB-L_mYM2}
\end{align}

For the modified version of the GF+FP term for the general Lorenz gauge:
\begin{align}
		\mathscr{L}_{\rm GF+FP}
		&= i {\boldsymbol \delta}_{\rm{B}} \bar{{\boldsymbol \delta}}_{\rm{B}} \left(\frac{1}{2} \mathscr{A}_{\mu} \cdot \mathscr{A}^{\mu} + \frac{\beta}{2} i \bar{\mathscr{C}} \cdot \mathscr{C} \right)
		+ \frac{\alpha}{2} \mathscr{N} \cdot \mathscr{N} ,
%\nonumber\\
%&= i {\boldsymbol \delta}' \bar{{\boldsymbol \delta}}' \left(\frac{1}{2} \mathscr{A}_{\mu} \cdot \mathscr{A}^{\mu} + \frac{\beta}{2} i \bar{\mathscr{C}} \cdot \mathscr{C} \right)
%- \frac{\beta}{2} M^2 i \bar{\mathscr{C}} \cdot \mathscr{C} + \frac{\alpha}{2} \mathscr{N} \cdot \mathscr{N} ,
\end{align}
its change under the generalized local gauge transformation is calculated as  
\begin{align}
		\delta^\omega \mathscr{L}_{\rm GF+FP}
		&= i {\boldsymbol \delta}_{\rm{B}}  \bar{{\boldsymbol \delta}}_{\rm{B}} \delta^\omega \left(\frac{1}{2} \mathscr{A}_{\mu} \cdot \mathscr{A}^{\mu}
		+ \frac{\beta}{2} i \bar{\mathscr{C}} \cdot \mathscr{C} \right) \nonumber\\
		&= i {\boldsymbol \delta}_{\rm{B}}  \bar{{\boldsymbol \delta}}_{\rm{B}}  \left( \mathscr{A}_{\mu} \cdot \delta^\omega \mathscr{A}^{\mu}
		+ \frac{\beta}{2} \delta^\omega (i \bar{\mathscr{C}} \cdot \mathscr{C}) \right) \nonumber\\
		&= i {\boldsymbol \delta}_{\rm{B}}  \bar{{\boldsymbol \delta}}_{\rm{B}} \mathscr{A}_{\mu} \cdot \partial^{\mu} \omega .
	\label{SSB-L_mYM}
\end{align}
%The change of the mass term reads
%\begin{align}
% \delta^\omega \mathscr{L}_m
% &= \delta^\omega \left(\frac{1}{2}M^2 \mathscr{A}_{\mu} \cdot \mathscr{A}^{\mu} + \beta M^2 i \bar{\mathscr{C}} \cdot \mathscr{C} \right) %\nonumber\\
% &= M^2 \mathscr{A}_{\mu} \cdot \delta^\omega \mathscr{A}^{\mu} + \beta M^2 \delta^H (i \bar{\mathscr{C}} \cdot \mathscr{C}) \nonumber\\
% &= M^2 \mathscr{A}_{\mu} \cdot \partial^{\mu} \omega .
%\end{align}
Thus we obtain
\begin{align}
 \delta^\omega \mathscr{L}(x) %^{\rm tot}_{\rm YM}(x)
=
%[M^2 \mathscr{A}^{\mu} +
 i {\boldsymbol \delta}_{\rm{B}} \bar{{\boldsymbol \delta}}_{\rm{B}}  \mathscr{A}^{\mu} (x) \cdot \partial_{\mu} \omega (x) .
	\label{SSB-L_tot}
\end{align}

	Equating (\ref{SSB-id2}) and (\ref{SSB-L_tot}), and then   comparing the terms proportional to $\omega$, $\partial^{\mu} \omega$ and $\partial^{\mu} \partial^{\nu} \omega$, we obtain 	
\begin{align}
  g \partial_{\mu} \mathscr{J}^{\mu}(x) 
	+ \frac{\delta S}{\delta \Phi}  \cdot gG =& 0 ,
		\nonumber\\
		    \frac{\delta S}{\delta \mathscr{A}_{\mu}(x)}  + \partial_{\nu} \mathscr{F}^{\mu \nu} + g \mathscr{J}^{\mu}(x) 
=& 
%M^2 \mathscr{A}^{\mu} + 
i {\boldsymbol \delta}_{\rm{B}} \bar{{\boldsymbol \delta}}_{\rm{B}}  \mathscr{A}^{\mu} .
 \end{align}
%where we have taken into account
%\begin{align}
%\frac{\partial \mathscr{L}^{\rm tot}_{m \rm YM}}{\partial (\partial_{\nu} \mathscr{A}_{\mu})}
%= \mathscr{F}^{\mu \nu} = - \mathscr{F}^{\nu \mu} .
%\end{align}
%By comparing the terms proportional to $\omega$, $\partial^{\mu} \omega$ and $\partial^{\mu} \partial^{\nu} \omega$, in both sides of (\ref{SSB-L_tot}) and (\ref{SSB-id}),   

The color current conservation holds:
\begin{align}
		\partial_{\mu} \mathscr{J}^{\mu} = 0 ,
\end{align}
and the field equation for $\mathscr{A}_{\mu}$ is written in the Maxwell-like form: 
\begin{align}
 \partial_{\nu} \mathscr{F}^{\mu \nu} + g \mathscr{J}^{\mu}
= 
%M^2 \mathscr{A}^{\mu} +
 i {\boldsymbol \delta}_{\rm{B}} \bar{{\boldsymbol \delta}}_{\rm{B}}  \mathscr{A}^{\mu} ,
\end{align}
where we have used antisymmetry of the field strength $\mathscr{F}^{\mu \nu}(x)=-\mathscr{F}^{\nu\mu}(x)$ and symmetry $\partial_{\mu} \partial_{\nu} \omega(x)=\partial_{\nu} \partial_{\mu} \omega(x)$ to yield $\mathscr{F}^{\mu \nu}(x) \cdot \partial_{\mu} \partial_{\nu} \omega(x)=0$.
[
This is not the case if there are line singularities in $\Lambda(x)$:
$
\mathscr{F}^{\mu \nu}(x) \cdot \partial_{\mu} \partial_{\nu} \omega(x)
= \frac12 \mathscr{F}^{\mu \nu}(x) \cdot [\partial_{\mu} , \partial_{\nu} ] \omega(x) \not=0.
$
]

By taking the divergence of both sides,  we obtain the consistent relation
\begin{align}
 \partial_{\mu} (%M^2 \mathscr{A}^{\mu} + 
 i {\boldsymbol \delta}_{\rm{B}} \bar{{\boldsymbol \delta}}_{\rm{B}}  \mathscr{A}^{\mu} ) = 0 ,
\end{align}
since $\partial_{\mu} \partial_{\nu} \mathscr{F}^{\mu \nu} = - \partial_{\mu} \partial_{\nu} \mathscr{F}^{\nu \mu} =0$
and $\partial_{\mu} \mathscr{J}^{\mu} = 0$.

%By taking into account
%\begin{align}
%\frac{\partial \mathscr{L} }{\partial (\partial_{\nu} \mathscr{A}_{\mu})}
%= \mathscr{F}^{\mu \nu} = - \mathscr{F}^{\nu \mu}  ,
%\end{align}
%Then the on mass-shell relation reads
%\begin{align}
% \delta^\omega \mathscr{L}(x) =& 
%  \left[   \partial_{\nu} \mathscr{F}^{\mu \nu}(x) + g \mathscr{J}^{\mu}(x) \right] \cdot \partial_{\mu} \omega(x) ,
%\end{align}

\subsection{MA gauge}

The MA gauge is defined by the GF+FP term of the form:
\begin{align}
\mathscr{L}_{\rm GF+FP}^{\rm MAG} = -i \bm{\delta}_{\rm{B}} \biggl[ \bar{C}^{a} \left( ( D^{\mu}[a] A_{\mu} )^{a} + \frac{\alpha}{2} B^{a} \right) \biggr] .
\label{C25-GF+FP-MAG}
\end{align}
For $G=SU(N)$,  the  GF+FP term of the MA gauge has the explicit form
\begin{align}
\mathscr{L}_{\rm GF+FP}^{\rm MAG} 
%=& B^{a} F^{a} +  \frac{\alpha}{2} B^{a} B^{a} + i \bar{C}^{a} D_{\mu}^{ab} [a] D^{\mu bc} [a] C^{c} \nonumber\\&
%+ i \bar{C}^{a} D_{\mu}^{ab} [a] ( g f^{bck} A^{\mu c} C^{k} + g f^{bcd} A^{\mu c} C^{d} ) 
%\nonumber\\&
%+ i \bar{C}^{a} g f^{akb} ( \partial_{\mu} C^{k} + g f^{kcd} A_{\mu}^{c} C^{d} ) A^{\mu b} \nonumber\\
=&  B^{a} F^{a} + \frac{\alpha}{2} B^{a} B^{a} 
%\nonumber\\&
+ i \bar{C}^{a} D_{\mu}^{ab} [a] D^{\mu bc} [a]  C^{c} + i g^{2} f^{kba} f^{kcd} \bar{C}^{a} C^{d} A_{\mu}^{b} A^{\mu c} \nonumber\\
&+ i \bar{C}^{a} D_{\mu}^{ab} [a] ( g f^{bcd} A^{\mu c} C^{d} ) 
%\nonumber\\&
+ i \bar{C}^{a} D_{\mu}^{ab} [a] ( g f^{bck} A^{\mu c} C^{k} ) + i \bar{C}^{a} g f^{akb} \partial_{\mu} C^{k} A^{\mu b} .
\end{align}
For $G=SU(2)$, the MA gauge has the relatively simple GF+FP term: 
\begin{align}
\mathscr{L}_{\rm GF+FP}^{\rm MAG} =&
B^{a} D_{\mu}[a]^{ab} A^{\mu b} + \frac{\alpha}{2} B^{a} B^{a} 
%\nonumber\\&
+ i \bar{C}^{a} D_{\mu}[a]^{ac} D^{\mu}[a]^{cb} C^{b} 
\nonumber\\&
+  g^{2} \varepsilon^{ad} \varepsilon^{bc} i\bar{C}^{a} C^{b} A^{\mu c} A_{\mu}^{d} 
+ i \bar{C}^{a} g \varepsilon^{ab} ( D^{\mu}[a]^{bc} A_{\mu}^{c} ) C^{3} .
\label{C25-MAG-SU2}
\end{align}
where $D_{\mu}[a]^{ab}= \delta^{ab} \partial_{\mu}   -  g \epsilon^{abj} a_{\mu}^{j}$.

%\footnote{
%H. Min, T. Lee, and P.Y. Pac,
%Renormalization Of Yang-mills Theory In The Abelian Gauge,
%Phys.Rev. D\textbf{32}, 440--449 (1985).
%}
%K. -I. Kondo,
%Yang-Mills theory as a deformation of topological field theory, dimensional reduction and quark confinement, 
%CHIBA-EP-103 
%hep-th/9801024, 
%Phys. Rev. D\textbf{58},  105019 (1998). 
However, the MA gauge given above is not renormalizable. %e \cite{MLP85}.  
Therefore, a renormalizable MA gauge is proposed in \cite{KondoII,Kondo01}, which we call the modified MA gauge. 
The modified MA gauge has the GF+FP term:
\begin{align}
		\mathscr{L}_{\rm GF+FP}^{\rm mMAG}
		&= i {\boldsymbol \delta}_{\rm{B}} \bar{{\boldsymbol \delta}}_{\rm{B}} \left(\frac{1}{2}  {A}_{\mu}^{a}  {A}^{\mu}{}^{a} + \frac{\alpha}{2} i \bar{{C}}^a   {C}^a \right)  ,
\end{align}
where the summation over only the off-diagonal components with the index $a$ is understood. 
The difference between the MA gauge and the modified MA gauge for $SU(2)$ is 
\begin{align}
\mathscr{L}_{\rm GF+FP}^{{\rm mMAG} } 
=&
-i \bm{\delta}_{\rm{B}} \biggl[ \bar{C}^{a} \left( (D_{\mu}[a] A^{\mu})^{a} + \frac{\alpha}{2} B^{a} \right) 
%\nonumber\\&
- i \frac{\alpha}{2} g f^{abk} \bar{C}^{a} \bar{C}^{b} C^{k} - i \frac{\alpha}{4} g f^{abc} C^{a} \bar{C}^{b} \bar{C}^{c} \biggr] 
\nonumber\\
=& \mathscr{L}_{\rm GF+FP}^{{\rm  MAG} } 
 - \alpha g \epsilon^{ab} B^{a}  i \bar{C}^{b}C^{3} + \frac{\alpha}{4} g^{2} \epsilon^{ab} \epsilon^{cd} \bar{C}^{a} \bar{C}^{b} C^{c} C^{d}  
,
\label{C25-mMAG2}
\end{align}
where we have used $f^{ajk}=0$ for $SU(N)$ and $f^{ajk}=0, f^{abc}=0$ for $SU(2)$. 
The ghost self-interaction term makes the modified MA gauge renormalizable.

We calculate the change of the GF+FP term under the generalized local gauge transformation as  
\begin{align}
  \delta^\omega \mathscr{L}_{\rm GF+FP}^{\rm mMAG}
 &= i {\boldsymbol \delta}_{\rm{B}}  \bar{{\boldsymbol \delta}}_{\rm{B}} \delta^\omega \left(\frac{1}{2}   {A}^{\mu}{}^{a} {A}_{\mu}^{a} 
%+ \frac{\alpha}{2} i \bar{{C}}^a   {C}^a 
\right)  = i {\boldsymbol \delta}_{\rm{B}}  \bar{{\boldsymbol \delta}}_{\rm{B}}  \left(    {A}^{\mu}{}^{a} \delta^\omega {A}_{\mu}^{a} 
%+ \frac{\beta}{2} \delta^\omega (i \bar{\mathscr{C}} \cdot \mathscr{C}) 
\right) \nonumber\\
 &=  i {\boldsymbol \delta}_{\rm{B}}  \bar{{\boldsymbol \delta}}_{\rm{B}}  \left(     {A}_{\mu}^{a} \partial^\mu \omega^a + g f^{bja} a^{\mu}{}^j A_\mu^b \omega^a 
%+ \frac{\beta}{2} \delta^\omega (i \bar{\mathscr{C}} \cdot \mathscr{C}) 
\right) 
\nonumber\\
 &=  i {\boldsymbol \delta}_{\rm{B}}  \bar{{\boldsymbol \delta}}_{\rm{B}} ({A}_{\mu}^{a} ) \partial^\mu \omega^a
+ i {\boldsymbol \delta}_{\rm{B}}  \bar{{\boldsymbol \delta}}_{\rm{B}} (g f^{bja} a^{\mu}{}^j A_\mu^b ) \omega^a
%+ \frac{\beta}{2} \delta^\omega (i \bar{\mathscr{C}} \cdot \mathscr{C}) 
,
\label{SSB-L_mYM_MAG}
\end{align}
where we have used $A_\mu^a A_\mu^b f^{abE}=0$.

The Lorenz gauge for the diagonal part is given by
\begin{align}
\mathscr{L}_{\rm GF+FP}^{\rm diag.} 
= -i \bm{\delta}_{\rm{B}} \biggl[ \bar{C}^{j} \left( \partial^{\mu} a_{\mu}^{j} + \frac{\beta}{2} B^{j} \right) \biggr] .
\end{align}
The change  under the generalized local gauge transformation is   
\begin{align}
\delta^\omega \mathscr{L}_{\rm GF+FP}^{\rm diag.}  
&= -i \bm{\delta}_{\rm{B}} \delta^\omega  \biggl[ \bar{C}^{j}  \left( \partial^{\mu} a_{\mu}^{j} 
%+ \frac{\beta}{2} B^{j} 
\right) \biggr] 
\nonumber\\
 &= i \bm{\delta}_{\rm{B}}(\partial_\mu \bar c^j ) \partial^\mu \omega^j 
-i \bm{\delta}_{\rm{B}}(\bar c^b \partial^\mu a_\mu^j ) gf^{jba} \omega^a 
+ i \bm{\delta}_{\rm{B}}(\partial^\mu \bar c^j A_\mu^b) gf^{jba} \omega^a  .
\end{align}

We find that the field equation for $\mathscr{A}_{\mu}$ is written in the Maxwell-like form also in the MA gauge by observing the $\partial^\mu \omega$ dependent terms. 
In the MA gauge, however, the explicit forms are different between the diagonal and off-diagonal components: 
The off-diagonal components obey
\begin{align}
 \partial^{\nu} {F}_{\mu \nu}^a + g {J}_{\mu}^a
=  i {\boldsymbol \delta}_{\rm{B}} \bar{{\boldsymbol \delta}}_{\rm{B}} {A}_{\mu}^a 
=  i {\boldsymbol \delta}_{\rm{B}} (\mathscr{D}_{\mu}[\mathscr{A}]\bar{\mathscr{C}})^a ,
\label{field-eq-off}
\end{align}
while the diagonal components obey
\begin{align}
 \partial^{\nu} {f}_{\mu \nu}^j + g {J}_{\mu}^j
=  i {\boldsymbol \delta}_{\rm{B}} (\partial_\mu \bar c^j)  = - \partial_\mu b^j .
\label{field-eq-diag}
\end{align}
Since $\delta^\omega \mathscr{L}_{\rm GF+FP}^{\rm mMAG}$ and $\delta^\omega \mathscr{L}_{\rm GF+FP}^{\rm diag.}$ contain the $\omega$ dependent terms in contrast to the Lorenz gauge, the color current conservation does not  hold, $\partial_{\mu} \mathscr{J}^{\mu} \not= 0$, in the MA gauge. 
To obtain the non-conservation term, it is simpler to use the field equations rather than using $\delta^\omega \mathscr{L}_{\rm GF+FP}^{\rm mMAG}$  and $\delta^\omega \mathscr{L}_{\rm GF+FP}^{\rm diag.}$.
By taking the divergence of (\ref{field-eq-off}) and (\ref{field-eq-diag}), we find that the off-diagonal  global current obey
\begin{align}
g \partial^\mu {J}_{\mu}^a
%=  i {\boldsymbol \delta}_{\rm{B}} \bar{{\boldsymbol \delta}}_{\rm{B}} g \partial^\mu {A}_{\mu}^a  
=  i {\boldsymbol \delta}_{\rm{B}} \partial^\mu (\mathscr{D}_{\mu}[\mathscr{A}]\bar{\mathscr{C}})^a ,
\label{eq:global_current_div_off}
\end{align}
while the diagonal  global current  obey
\begin{align}
g \partial^\mu {j}_{\mu}^j
=  i  {\boldsymbol \delta}_{\rm{B}} (\partial^\mu \partial_\mu \bar c^j)  = -  \partial^\mu \partial_\mu b^j .
\label{eq:global_current_div}
\end{align}
The diagonal part is further simplified as follows. 
The field equation for $B^a$ reads
\begin{align}
  0 = \frac{\delta S}{\delta B^a} = \mathscr{D}^\mu [a]^{a b} A_\mu^b + \alpha B^a .
\end{align}
The field equation for $\bar{c}^j$ is obtained by using the left derivative as
\begin{align}
  0 = \frac{\delta S}{\delta c^j} = &i g (\mathscr{D}^\mu [a]^{b a} \bar{C}^a) f^{b c j} A_\mu^c + i g \partial_\mu (\bar{C}^b f^{b j c} A^{\mu c}) - i \partial_\mu \partial^\mu \bar{c}^j \nonumber\\
  = &i g f^{j b c} (\mathscr{D}^\mu [a]^{b a} \bar{C}^a) A_\mu^c - i g f^{j b c} \partial_\mu (\bar{C}^b A^{\mu c}) - i \partial_\mu \partial^\mu \bar{c}^j \nonumber\\
  = &i g f^{j b c} g f^{b k a} a^{\mu k} \bar{C}^a A_\mu^c + i g f^{j b c} \bar{C}^c \partial_\mu A^{\mu b} - i \partial_\mu \partial^\mu \bar{c}^j \nonumber\\
  = &i g f^{j a b} g f^{k b c} a^{\mu k} \bar{C}^a A_\mu^c + i g f^{j b c} \bar{C}^c \partial_\mu A^{\mu b} - i \partial_\mu \partial^\mu \bar{c}^j \nonumber\\
  = &- i g f^{j a b} \bar{C}^a g f^{b k c} a^{\mu k} A_\mu^c - i g f^{j a b} \bar{C}^a \partial_\mu A^{\mu b} - i \partial_\mu \partial^\mu \bar{c}^j \nonumber\\
  = &- i g f^{j a b} \bar{C}^a \mathscr{D}^\mu [a]^{b c} A_\mu^c - i \partial_\mu \partial^\mu \bar{c}^j \nonumber\\
  = &- i \partial_\mu \partial^\mu \bar{c}^j + \alpha i g f^{j a b} \bar{C}^a B^b ,
\end{align}
where we have used the Jacobi identity $f^{k a b} f^{j b c} = f^{k a B} f^{j B c} = - f^{a j B} f^{k B c} - f^{j k B} f^{a B c} = - f^{a j b} f^{k b c}$ in the forth equality and  the field equation for $B^a$ in the last equality.
Therefore, the divergence of the diagonal global current \eqref{eq:global_current_div} vanishes for any $\alpha$:
\begin{align}
  g \partial^\mu j_\mu^j = i \bm{\delta}_B (\partial^\mu \partial_\mu \bar{c}^j) =  \alpha i g f^{j a b} \bm{\delta}_B (\bar{C}^a B^b)  
  =  \alpha i g f^{j a b} i B^a B^b  
  =  0 , %\ (\forall \alpha)
\end{align}
where we have used $\bm{\delta}_B \bar{C}^a = i B^a$, $\bm{\delta}_B B^b = 0$ and $\ f^{j a b} = - f^{j b a}$.
The diagonal global current is conserved in the MA gauge, in contrast to the off-diagonal current which is not conserved.
Thus, the diagonal local current has the divergence
\begin{align}
  \partial^\mu j_\mu^{\omega} = i \bm{\delta}_B \partial_\mu \bar{c}^j \partial^\mu \omega^j = - \partial_\mu b^j \partial^\mu \omega^j  .
  \label{eq:MA-current-divergence-diagonal}
\end{align}
Finally, we have the local gauge transformation of the Lagrangian given by
\begin{align}
		\delta^\omega \mathscr{L}  
&= 
 g \partial_{\mu} \mathscr{J}^{\mu}  \cdot \omega 
+ \left[    \partial_{\nu} \mathscr{F}^{\mu \nu}   + g \mathscr{J}^{\mu}  \right] \cdot \partial_{\mu} \omega 
\nonumber\\  
&= 
 g \partial^\mu {J}_{\mu}^a \omega^a + g \partial^\mu {J}_{\mu}^j \omega^j 
+ \left[ \partial^{\nu} {F}_{\mu \nu}^a + g {J}_{\mu}^a \right]  \partial_{\mu} \omega^a
+ \left[ \partial^{\nu} {f}_{\mu \nu}^j + g {J}_{\mu}^j \right]  \partial_{\mu} \omega^j
\nonumber\\  
%- \mathscr{F}^{\mu \nu}(x) \cdot \partial_{\mu} \partial_{\nu} \omega(x) 
&= i {\boldsymbol \delta}_{\rm{B}} \partial^\mu (\mathscr{D}_{\mu}[\mathscr{A}]\bar{\mathscr{C}})^a \omega^a  
%-  \partial^\mu \partial_\mu b^j \omega^j  
+ i {\boldsymbol \delta}_{\rm{B}} (\mathscr{D}_{\mu}[\mathscr{A}]\bar{\mathscr{C}})^a \partial^\mu \omega^a   
- \partial_\mu b^j \partial^\mu \omega^j  .
		\label{SSB-MAG1}
\end{align}

\begin{comment}
Some terms cancel as follows. 
\begin{align}
 & \bar{{\boldsymbol \delta}}_{\rm{B}} (g f^{bja} a^{\mu}{}^j A_\mu^b ) \omega^a
\nonumber\\   &=   g f^{bja} \bar{{\boldsymbol \delta}}_{\rm{B}} a^{\mu}{}^j A_\mu^b  \omega^a
   + g f^{bja} a^{\mu}{}^j \bar{{\boldsymbol \delta}}_{\rm{B}}  A_\mu^b   \omega^a  
\nonumber\\
 &= g f^{bja} (\partial^\mu \bar c^j + gf^{iac} A_\mu^{a} \bar c^{c} ) A_\mu^b  \omega^a
   + g f^{bja} a^{\mu}{}^j (\partial^\mu \bar c^b + gf^{bkc} a_\mu^{k} \bar c^{c} + gf^{bck} A_\mu^{c} \bar c^{k} + gf^{bcd} A_\mu^{c} \bar c^{d})   \omega^a  
\nonumber\\
 &= g f^{bja}  \partial^\mu \bar c^j  A_\mu^b  \omega^a
   + g f^{bja} a^{\mu}{}^j (\partial^\mu \bar c^b + gf^{bkc} a_\mu^{k} \bar c^{c} + gf^{bck} A_\mu^{c} \bar c^{k} + gf^{bcd} A_\mu^{c} \bar c^{d})   \omega^a  
  ,
\label{SSB-L_mYM_MAG}
\end{align}
\end{comment}

\section{Relations among 2-point functions}

The Slavnov-Taylor identity holds, 
\begin{align}
0=  \langle 0 | \bm{\delta}_{\rm{B}}[ {\rm T} \bar{\mathscr{C}}^A(x) \mathscr{A}_{\mu}^B(y)  ]   | 0 \rangle 
= - \langle 0 | {\rm T} \bar{\mathscr{C}}^A(x) ( \mathscr{D}_{\mu} \mathscr{C} )^B(y)   | 0 \rangle 
+ \langle 0 | {\rm T} i \mathscr{B}^A(x) \mathscr{A}_{\mu}^B(y)  | 0 \rangle ,
\end{align}
which yields
\begin{align}
  \langle 0 | {\rm T} \bar{\mathscr{C}}^A(x) ( \mathscr{D}_{\mu} \mathscr{C} )^B(y)   | 0 \rangle 
= \langle 0 | {\rm T} i \mathscr{B}^A(x) \mathscr{A}_{\mu}^B(y) | 0 \rangle .
\end{align}

We show that 
\begin{align}
  \langle 0 | {\rm T} \bar{\mathscr{C}}^A(x) \partial^\mu ( \mathscr{D}_{\mu} \mathscr{C} )^k(y)   | 0 \rangle 
= -  \delta^D(x-y) \delta^{Ak}  .
\end{align}

In the path-integral quantization, this relation is derived as 
\begin{align}
  \langle 0| \bar{\mathscr{C}}^A(x) \partial^\mu ( \mathscr{D}_{\mu} \mathscr{C} )^k(y) |0 \rangle 
%\nonumber\\
 =& \int d\mu(\Phi)  e^{iS_{\rm YM}^{\rm tot}}  \bar{\mathscr{C}}^A(x) \partial^\mu ( \mathscr{D}_{\mu} \mathscr{C} )^k(y)
\nonumber\\
 =& \int d\mu(\Phi)  \bar{\mathscr{C}}^A(x)  \frac{\delta e^{iS_{\rm YM}^{\rm tot}}}{\delta (-\bar{{c}}^k(y))} 
\nonumber\\
 =& \int d\mu(\Phi) \frac{\delta}{\delta\bar{{c}}^k(y)}\left[ \bar{\mathscr{C}}^A(x) e^{iS_{\rm YM}^{\rm tot}} \right]
 - \int d\mu e^{iS_{\rm YM}^{\rm tot}}\frac{\delta \bar{\mathscr{C}}^A(x)}{\delta \bar{{c}}^k(y)}
\nonumber\\
 =& - \delta ^{Ak}\delta ^D(x-y)
  ,
\end{align}
where we have used a fact that the integration of the derivative with respect to the fundamental field $\Phi _I(x)$ is identically zero, which is called the Schwinger-Dyson equation:
%
\begin{comment}:
\footnote{
This relation is called the \textbf{Schwinger-Dyson equation}, which is a consequence of 
$
  \int_{a}^{b} d\phi \frac{d f}{d \phi}
  = f(b)-f(a) =0 
$
where we have used $f(b)=f(a)$ at the boundary.
This relation follows also from the shift invariance of the measure, since 
$
  \int d\phi f(\phi) = \int d(\phi+a) f(\phi+a) = \int d\phi f(\phi+a) 
= \int d\phi f(\phi)+a \int d\phi \frac{df(\phi)}{d\phi} + O(a^2)  
$
holds for arbitrary $a$.
}
\end{comment}
%
\begin{align}
 \int d\mu(\Phi) \frac{\delta}{\delta\Phi _I(x)}\left[\cdots\right] \equiv 0 ,
\end{align}
and 
$\int d\mu e^{iS_{\rm YM}^{\rm tot}}=1$.
We have
\begin{align}
  \partial^\mu_y  \langle 0 | {\rm T} \bar{\mathscr{C}}^A(x)  ( \mathscr{D}_{\mu} \mathscr{C} )^k(y)   | 0 \rangle 
= - \delta^D(x-y)  \delta^{Ak}  .
\end{align}
Thus we obtain
\begin{align}
\langle 0 | {\rm T} i \mathscr{B}^A(x) \mathscr{A}_{\mu}^k(y) | 0 \rangle 
= \langle 0 | {\rm T} \bar{\mathscr{C}}^A(x)  ( \mathscr{D}_{\mu} \mathscr{C} )^k(y)   | 0 \rangle 
= - \frac{\partial_\mu^y}{\partial_y^2}  \delta^D(x-y) \delta^{Ak}  
= \frac{\partial_\mu^x}{\partial_x^2}  \delta^D(x-y) \delta^{Ak}  .
\end{align}
This relation is also obtained in the canonical quantization. 

We show that the following 2-point function is written in the transverse form:
\begin{align}
\langle 0 | {\rm T} (g \mathscr{A}_{\nu} \times \bar{\mathscr{C}})^A(x) (\mathscr{D}_{\mu}[\mathscr{A}] \mathscr{C})^k(y)  | 0 \rangle
= \left(g_{\mu\nu}- \frac{\partial_{\mu}^x \partial_{\nu}^x}{\partial_x^2} \right) v^{Ak}(x-y) ,
\label{CCF-tra-MAG}
\end{align}
where $v^{AB}$ is the  Kugo-Ojima (KO) function in the MA gauge.
The transversality (\ref{CCF-tra-MAG}) follows from the fact that the Schwinger-Dyson equation 
\begin{align}
0 &= \int d\mu \frac{\delta}{\delta \bar{{c}}^k(y)}
\left[ (g \mathscr{A}_{\nu} \times \bar{\mathscr{C}})^A(x)  e^{i S^{\rm tot}_{\rm YM}}  \right] \nonumber\\
&= \int d\mu\  \Big[ g f^{ABk} \mathscr{A}_{\nu}^B(x) \delta^D(x-y) %\nonumber\\ &\quad\quad\quad\quad
- (g \mathscr{A}_{\nu} \times \bar{\mathscr{C}})^A(x)  \partial^{\mu} (\mathscr{D}_{\mu} \mathscr{C})^k(y) \Big] e^{i S^{\rm tot}_{\rm YM}}  ,
\end{align}
 yields
\begin{align}
 \langle 0 | {\rm T} (g \mathscr{A}_{\nu} \times \bar{\mathscr{C}})^A(x) \partial_y^{\mu} (\mathscr{D}_{\mu} \mathscr{C})^k(y)  | 0 \rangle %\nonumber\\
 = gf^{ABk} \langle 0 | \mathscr{A}_{\nu}^B(x) | 0 \rangle \delta^D(x-y) = 0 ,
\end{align}
where we have used $\langle 0 | \mathscr{A}_{\nu}^B(x) | 0 \rangle=0$ which follows from the Lorentz invariance of the vacuum $| 0 \rangle$
and the Lorentz covariance of $\mathscr{A}_{\mu}(0)$.

\section{Topological configurations}

\subsection{D=3}

We show that the Dirac magnetic monopole and the Wu-Yang magnetic monopole are mutually  related by the gauge transformation. 
In other words, the Wu-Yang magnetic monopole are gauge equivalent to the Dirac magnetic monopole. 
We start from the Yang-Mills field of the Abelian type:
\begin{align}
  {A}_\mu^1(x) = 0 , \ {A}_\mu^2(x) = 0 , \ {A}_\mu^3(x) \not= 0 \Rightarrow \mathscr{A}_\mu(x) := \mathscr{A}_\mu^A(x) T_A = A_\mu^3(x) T_3 .
\end{align}
The gauge field $\mathscr{A}_\mu(x)$ is transformed into $\mathscr{A}_\mu^\prime(x)$ by the matrix $U (x) \in G$ of the gauge group $G$ as
\begin{align}
  \mathscr{A}_\mu^\prime(x)  =& U(x) \mathscr{A}_\mu(x) U^{- 1}(x) + i g^{- 1} U(x) \partial_\mu U^{- 1}(x)  
\nonumber\\  
  =& A_\mu^3(x) U(x) T_3 U^{- 1}(x) + i g^{- 1} U(x) \partial_\mu U^{- 1} (x) .
  \label{eq:gauge_transformation_by_U}
\end{align}
In general, the matrix $U \in G=SU (2)$ can be represented by the three Euler angles $\alpha, \beta , \gamma$ as 
\begin{align}
  U (x) = &e^{- i \alpha (x) \sigma_3 / 2} e^{- i \beta (x) \sigma_2 / 2} e^{- i \gamma (x) \sigma_3 / 2} \in SU(2)
  \nonumber\\
%= &
% \begin{pmatrix}
%  e^{\frac{i}{2}[- \alpha (x) - \gamma (x)]} \cos \frac{\beta (x)}{2} & - e^{\frac{i}{2}[- \alpha (x) + \gamma (x)]} \sin \frac{\beta (x)}{2} \\
%  e^{\frac{i}{2}[\alpha (x) - \gamma (x)]} \sin \frac{\beta (x)}{2} & e^{\frac{i}{2}[\alpha (x) + \gamma (x)]} \cos \frac{\beta (x)}{2}
% \end{pmatrix}
%\nonumber\\
%= &\cos \left(\frac{\alpha (x) + \gamma (x)}{2}\right) \cos \frac{\beta (x)}{2} I - i \sin \left(\frac{\gamma (x) - \alpha (x)}{2}\right) \sin \frac{\beta (x)}{2} \sigma_1 \nonumber\\
% &- i \cos \left(\frac{\gamma (x) - \alpha (x)}{2}\right) \sin \frac{\beta (x)}{2} \sigma_2 - i \sin \left(\frac{\alpha (x) + \gamma (x)}{2}\right) \cos \frac{\beta (x)}{2} \sigma_3
%\nonumber\\
  &\alpha (x) \in [0 , 2 \pi) , \ \beta (x) \in [0 , \pi] , \gamma (x) \in [0 , 2 \pi) .
\end{align}
Then the first term of 
\eqref{eq:gauge_transformation_by_U} is calculated as
\begin{align}
   A_\mu^3(x) U(x) T_3 U^{- 1}(x) 
  =  A_\mu^3(x) [\cos \alpha (x) \sin \beta (x) T_1 + \sin \alpha (x) \sin \beta (x) T_2 + \cos \beta (x) T_3] ,
\end{align}
and the second term of \eqref{eq:gauge_transformation_by_U} is 
\begin{align}
   iU(x) \partial_\mu U^{- 1}(x) \
%\nonumber\\
  = &- \partial_\mu \gamma (x) \cos \alpha (x) \sin \beta (x) T_1 + \partial_\mu \beta (x) \sin \alpha (x) T_1
\nonumber\\
   &- \partial_\mu \gamma (x) \sin \alpha (x) \sin \beta (x) T_2 - \partial_\mu \beta (x) \cos \alpha (x) T_2
\nonumber\\
&- \partial_\mu \alpha (x) T_3 - \partial_\mu \gamma (x) \cos \beta (x) T_3 ,
\end{align}
where we have used the relations: 
$\sigma_k^2=\bm{1} \ (k = 1 , 2 , 3)$, 
$\sigma_2 \sigma_3 =i \sigma_1 = - \sigma_3 \sigma_2$, 
$ \sigma_3 \sigma_1=i \sigma_2 = - \sigma_1 \sigma_3$, and  
$\sigma_1 \sigma_2 =i \sigma_3 = - \sigma_2 \sigma_1$.
Therefore, the gauge transformation \eqref{eq:gauge_transformation_by_U} is given by
\begin{align}
  \mathscr{A}^\prime_\mu(x) = &[A_\mu^3(x) \cos \alpha (x) \sin \beta (x) - g^{- 1} \partial_\mu \gamma (x) \cos \alpha (x) \sin \beta (x) + g^{- 1} \partial_\mu \beta (x) \sin \alpha (x)] T_1 \nonumber\\
  &+ [A_\mu^3(x) \sin \alpha (x) \sin \beta (x) - g^{- 1} \partial_\mu \gamma (x) \sin \alpha (x) \sin \beta (x) - g^{- 1} \partial_\mu \beta (x) \cos \alpha (x)] T_2 \nonumber\\
  &+ [A_\mu^3(x) \cos \beta (x) - g^{- 1} \partial_\mu \gamma (x) \cos \beta (x) - g^{- 1} \partial_\mu \alpha (x) ] T_3 .
\end{align}
Therefore, the gauge field of the Wu-Yang magnetic monopole,
\begin{align}
  \mathscr{A}_\mu^\prime(x) = C \epsilon_{A \mu \nu} \frac{x_\nu}{|x|^2} T_A ,
\end{align}
is realized if and only if $A_\mu^3(x)$ simultaneously satisfy the three equations:
\begin{align}
  A_\mu^3(x) - g^{- 1} \partial_\mu \gamma (x)  &= - g^{- 1} \partial_\mu \beta (x) \frac{\tan \alpha (x)}{\sin \beta (x)} + C \frac{\epsilon_{1 \mu \nu} x_\nu}{|x|^2 \cos \alpha (x) \sin \beta (x)} , \nonumber\\
 A_\mu^3(x) - g^{- 1} \partial_\mu \gamma (x)  &=  g^{- 1} \partial_\mu \beta (x) \frac{\cot \alpha (x)}{\sin \beta (x)} + C \frac{\epsilon_{2 \mu \nu} x_\nu}{|x|^2 \sin \alpha (x) \sin \beta (x)} , \nonumber\\
  A_\mu^3(x) - g^{- 1} \partial_\mu \gamma (x)  &= g^{- 1} \partial_\mu \alpha (x) \frac{1}{\cos \beta (x)} + C \frac{\epsilon_{3 \mu \nu} x_\nu}{|x|^2 \cos \beta (x)} .
  \label{eq:A_derived}
\end{align}
For this purpose, the three Euler angles $\alpha(x) , \beta(x) , \gamma(x)$ must be chosen appropriately so that the three equations become identical. 
Note that $\gamma(x)$ is arbitrary, since $g^{- 1} \partial_\mu \gamma (x)$
corresponds to the gauge transformation for the Abelian gauge field. 

Here we introduce the polar coordinate $(r , \theta , \varphi)$ for the three spatial dimensions. 
For concreteness, we adopt the Ansatz: $\beta(x) = \beta (\theta , \varphi) , \ \alpha(x)= \alpha (\varphi)$.
%First, we consider the $\bm{e}_\varphi$ component. 
The right-hand-side (RHS) of the third equation of \eqref{eq:A_derived} has only the $\bm{e}_\varphi$ component and the vanishing $\bm{e}_\theta$ component, since $\partial_\mu \alpha (\varphi)$ is parallel to $\bm{e}_\varphi$, i.e., 
$\partial_\mu \alpha (\varphi) \parallel \bm{e}_\varphi$ and 
\begin{align}
  C \frac{\epsilon_{3 \mu \nu} x_\nu}{|x|^2 \cos \beta} = - \frac{C \sin \theta}{r \cos \beta}(- \sin \varphi , \cos \varphi , 0) = - \frac{C \sin \theta}{r \cos \beta} \bm{e}_\varphi .
\end{align}
In the RHS of the first and second equations of \eqref{eq:A_derived}, therefore, the $\bm{e}_\theta$ component  must vanish.  
By taking into account
\begin{align}
  C \frac{\epsilon_{1 \mu \nu} x_\nu}{|x|^2 \cos \alpha \sin \beta} &= \frac{C}{r \cos \alpha \sin \beta} (0 , \cos \theta , - \sin \theta \sin \varphi) \nonumber\\
  &= \frac{C (\sin \varphi \bm{e}_\theta + \cos \theta \cos \varphi \bm{e}_\varphi)}{r \cos \alpha \sin \beta}  , \\
  C \frac{\epsilon_{2 \mu \nu} x_\nu}{|x|^2 \sin \alpha \sin \beta} &= \frac{C}{r \sin \alpha \sin \beta} (- \cos \theta , 0 , \sin \theta \cos \varphi) \nonumber\\
  &= \frac{C (- \cos \varphi \bm{e}_\theta + \cos \theta \sin \varphi \bm{e}_\varphi)}{r \sin \alpha \sin \beta} ,
\end{align}
therefore, 
the RHS of the first equation of \eqref{eq:A_derived} yields
\begin{align}
   - g^{- 1} \frac{\partial \beta}{\partial \theta} \frac{\tan \alpha}{\sin \beta} + \frac{C \sin \varphi}{\cos \alpha \sin \beta} = 0 %\nonumber\\
  \Leftrightarrow \  g^{- 1} \frac{\partial \beta}{\partial \theta} \sin \alpha = C \sin \varphi ,
  \label{eq:angle_condition1}
\end{align}
and the RHS of the second equation of \eqref{eq:A_derived} yields
\begin{align}
   g^{- 1} \frac{\partial \beta}{\partial \theta} \frac{\cot \alpha}{\sin \beta} - C \frac{\cos \varphi}{\sin \alpha \sin \beta} = 0 %\nonumber\\
  \Leftrightarrow \  g^{- 1} \frac{\partial \beta}{\partial \theta} \cos \alpha = C \cos \varphi .
  \label{eq:angle_condition2}
\end{align}
By dividing both sides of \eqref{eq:angle_condition1} by \eqref{eq:angle_condition2}, we obtain 
\begin{align}
   \tan \alpha = \tan \varphi %\nonumber\\
  \ \Leftrightarrow \  \alpha = \varphi + n \pi \ (n \in \mathbb{Z}) .
  \label{eq:angle_condition3}
\end{align}
The two equations \eqref{eq:angle_condition1} and \eqref{eq:angle_condition2} show that $\beta$  depends on $\theta$ in such a way that 
\begin{align}
   \frac{\partial \beta}{\partial \theta} = \pm \frac{C}{g^{- 1}} =: \pm c_1 = \text{const.} %\nonumber\\
  \Leftrightarrow \  \beta(\theta, \varphi) = \pm c_1 \theta + c_0 (\varphi) ,
  \label{eq:angle_condition4}
\end{align}
where the signature is $+$ for $n = \text{even}$, and $-$ for $n = \text{odd}$.

For the $\bm{e}_\varphi$ component,  the three equations must hold simultaneously:
\begin{align}
  &- g^{- 1} \frac{\partial \beta}{\partial \varphi} \frac{\tan \alpha}{\sin \theta \sin \beta} + \frac{C \cos \theta \cos \varphi}{\cos \alpha \sin \beta} 
  = g^{- 1} \frac{\partial \beta}{\partial \varphi} \frac{\cot \alpha}{\sin \theta \sin \beta} + \frac{C \cos \theta \sin \varphi}{\sin \alpha \sin \beta} \nonumber\\
  = &g^{- 1} \frac{d \alpha}{d \varphi} \frac{1}{\sin \theta \cos \beta} - \frac{C \sin \theta}{\cos \beta} .
  \end{align}
After substituting $\alpha = \varphi + n \pi$, they are cast into 
\begin{align}  
%\Leftrightarrow \ 
&
- g^{- 1} \frac{\partial c_0}{\partial \varphi} \frac{\tan \varphi}{\sin \theta \sin \beta} \pm \frac{C \cos \theta}{\sin \beta} 
  = g^{- 1} \frac{\partial c_0}{\partial \varphi} \frac{\cot \varphi}{\sin \theta \sin \beta} \pm \frac{C \cos \theta}{\sin \beta} \nonumber\\
  = &g^{- 1} \frac{1}{\sin \theta \cos \beta} - \frac{C \sin \theta}{\cos \beta} .
  \label{eq:angle_condition5}
\end{align}
By comparing the first two equations, we find
\begin{align}
&- \frac{\partial c_0}{\partial \varphi} \tan \varphi = \frac{\partial c_0}{\partial \varphi} \cot \varphi \Leftrightarrow
\frac{\partial c_0}{\partial \varphi} \frac{1}{\sin \varphi \cos \varphi} = 0 .
\end{align}
For this equation to hold for arbitrary $\varphi$, $\frac{\partial c_0}{\partial \varphi}=0$ must hold, namely, $c_0$ does not depend on $\varphi$, implying $c_0 = \text{const.}$.
Therefore, \eqref{eq:angle_condition5} reduces to 
\begin{align}
  &\pm \frac{C \cos \theta}{\sin \beta} = g^{- 1} \frac{1}{\sin \theta \cos \beta} - \frac{C \sin \theta}{\cos \beta} ,
%\nonumber\\
%\Leftrightarrow \ &(g^{- 1} - C \sin^2 \theta) \sin \beta = \pm C \sin \theta \cos \beta \cos \theta
\end{align}
which is equivalent to 
\begin{align}
  c_1 =\frac{C}{g^{- 1}} =  \frac{\sin \beta}{\sin \theta (\sin \theta \sin \beta \pm \cos \theta \cos \beta)} 
= \pm \frac{\sin \beta}{\sin \theta \cos (\theta \mp \beta)} = \text{const.}
%\nonumber\\
% \Leftrightarrow \ c_1 &= \pm \frac{\sin \beta}{\sin \theta \cos (\theta \mp \beta)} = \text{const.}
  \label{eq:angle_condition6}
\end{align}
Thus, we have arrived at the conclusion: 
the gauge field of the Abelian type
\begin{align}
  A_\mu^3(x) = &g^{- 1} \partial_\mu \gamma (x) + g^{- 1} \partial_\mu \varphi \frac{1}{\cos \beta (x)} + c_1 g^{- 1} \frac{\epsilon_{3 \mu \nu} x_\nu}{|x|^2 \cos \beta (x)} \nonumber\\
  = &g^{- 1} \partial_\mu \gamma (x) + \frac{g^{- 1} (1 - c_1 \sin^2 \theta)}{r \sin \theta \cos \beta(x)} \bm{e}_\varphi , \quad \beta(x) = \pm c_1 \theta + c_0 ,
%\nonumber\\
% \biggl(c_1 = &\pm \frac{\sin \beta}{\sin \theta \cos (\theta \mp \beta)} = \text{const.}\biggl)
\end{align}
yields by the gauge transformation 
$U(x) = e^{-i(\varphi+n\pi) \sigma_3 / 2} e^{-i \beta(x) \sigma_2 / 2} e^{-i \gamma(x)\sigma_3 / 2}$
the non-Abelian gauge field of the Wu-Yang magnetic monopole
\begin{align}
  \mathscr{A}_\mu^\prime(x) = c_1 g^{- 1} \epsilon_{A \mu \nu} \frac{x_\nu}{|x|^2} T_A .
\end{align}
By choosing $c_1 = 1, c_0=m\pi$, i.e., 
\begin{align}
  \beta(x) = \pm \theta + m \pi \ (m \in \mathbb{Z}) ,
\end{align}
the Wu-Yang magnetic monopole of the unit magnetic charge,
\begin{align}
  \mathscr{A}_\mu^\prime(x) = g^{- 1} \epsilon_{A \mu \nu} \frac{x_\nu}{|x|^2} T_A,
\end{align}
is obtained by the gauge transformation $U(x) = e^{-i (\varphi+n\pi) \sigma_3 / 2} e^{-i (\pm \theta + m \pi) \sigma_2 / 2} e^{-i  \gamma(x)\sigma_3 / 2}$ from the gauge field of the Abelian type
\begin{align}
  A_\mu^3(x) 
%= &g^{- 1} \partial_\mu \gamma (x) + (- 1)^m g^{- 1} \partial_\mu \varphi \frac{1}{\cos \theta (x)} + (- 1)^m g^{- 1} \frac{\epsilon_{3 \mu \nu} x_\nu}{|x|^2 \cos \theta (x)} \nonumber\\
  =  g^{- 1} \partial_\mu \gamma (x) + (- 1)^m \frac{g^{- 1} \cos \theta}{r \sin \theta} \bm{e}_\varphi .
  \label{eq:A_derived1}
\end{align}
In particular, the choice $\gamma = \pm \varphi$ yields the Dirac magnetic monopole with a unit magnetic charge
%the three equations give the identical gauge field for the Dirac magentic monopole 
\begin{align}
 A_\mu^3(x) =  - \frac{g^{- 1}}{r} \frac{\pm 1- \cos \theta}{\sin \theta} \bm{e}_\varphi ,
\end{align}
which has singularities on either positive or negative $z$ axis. 
While the choice $\gamma = 0$ yields the Schwinger magnetic monopole
\begin{align}
 A_\mu^3(x) =    \frac{g^{- 1}}{r} \frac{\cos \theta}{\sin \theta} \bm{e}_\varphi ,
\end{align}
which has singularities on both positive and negative $z$ axis.

\subsection{D=4}

In the four-dimensional Euclidean space with the coordinate $x_\mu$, 
the self-dual equation for the Yang-Mills field strength is solved under the Ansatz 
\begin{align}
\mathscr{A}_\mu (x) = - g^{- 1} \bar{\Sigma}_{\mu \nu} \partial_\nu (\ln \phi (x)) ,
\end{align}
to give the Yang-Mills gauge field $\mathscr{A}_\mu (x)$ represented by 
\begin{align}
  \mathscr{A}_\mu (x) = 2 g^{- 1} \lambda^2 \bar{\Sigma}_{\mu \nu} \frac{(x - a_s)_\nu}{(x - a_s)^2 ((x - a_s)^2 + \lambda^2)} = 2 g^{- 1} \lambda^2 \bar{\Sigma}_{\mu \nu} \frac{y_\nu}{y^2 (y^2 + \lambda^2)}
  \label{eq:A_definition}
\end{align}
where we have introduced the shifted coordinate $y_\mu$ defined by $y_\mu = x_\mu - a_\mu$ and the symbols defined by 
\begin{align} 
\bar{\Sigma}_{\mu \nu} = \bar{\eta}_{k \mu \nu} \frac{\sigma_k}{2}, \quad 
  \bar{\eta}_{k \mu \nu} = - \bar{\eta}_{k \nu \mu} =
  \begin{cases}
    \epsilon_{k \mu \nu} &(\mu , \nu = 1 , 2 , 3) \\
    - \delta_{k \mu} &(\nu = 4)
  \end{cases} .
\end{align}
Here $\lambda$ is an arbitrary non-negative constant $\lambda \ge 0$ corresponding to the size of the instanton, 
This gauge field has the singularity at $y = 0 \Leftrightarrow \ x = a_s$, which is called the instanton in the singular gauge.

Introduce $U (x) \in SU (2)$ by
\begin{align}
  U (x) = \frac{1}{|x|}(x_4 \bm{1} + i x_k \sigma_k) ,
\end{align}
which has 
\begin{align}
  U^{- 1} (x) \partial_j U (x) 
%&= \frac{x_4 I - i x_i \sigma_i}{|x|}\left(\frac{i \sigma_j}{|x|} - \frac{x_j U (x)}{|x|^2}\right) \nonumber\\
%&= \frac{i x_4 \sigma_j + x_i (\delta_{i j} I + i \epsilon_{i j k} \sigma_k)}{|x|^2} - \frac{x_j I}{|x|^2} \nonumber\\
%&= i \frac{x_4 \sigma_j - \epsilon_{k j i} x_i \sigma_k}{|x|^2} \nonumber\\
%&
= - 2 i \bar{\Sigma}_{j \nu} \frac{x_\nu}{|x|^2} , \quad 
%\end{align}
%and
%\begin{align}
  U^{- 1} (x) \partial_4 U (x) 
%&= \frac{x_4 I - i x_i \sigma_i}{|x|} \left(\frac{I}{|x|} - \frac{x_4 U (x)}{|x|^2}\right) \nonumber\\
%&= \frac{x_4 I - i x_i \sigma_i}{|x|^2} - \frac{x_4 I}{|x|^2} \nonumber\\
%&= - i \frac{x_i \sigma_i}{|x|^2} \nonumber\\
%&= - i \frac{\delta_{i \nu} x_\nu \sigma_i}{|x|^2} \nonumber\\
%&
= - 2 i \bar{\Sigma}_{4 \nu} \frac{x_\nu}{|x|^2} .
\end{align}
Then, the gauge field in the singular gauge is rewritten into 
\begin{align}
  \mathscr{A}_\mu (x) = i g^{- 1} U^{- 1} (y) \partial_\mu U (y) \frac{\lambda^2}{y^2 + \lambda^2} .
\end{align}

We perform the gauge transformation by the matrix $U (y) \in SU (2)$ to obtain
\begin{align}
  \mathscr{A}_\mu^\prime (x) = &U (y) \mathscr{A}_\mu (x) U^{- 1} (y) + i g^{- 1} U (y) \partial_\mu U^{- 1} (y) \nonumber\\
  = &i g^{- 1} \partial_\mu U (y) U^{- 1} (y) \frac{\lambda^2}{y^2 + \lambda^2} + i g^{- 1} U (y) \partial_\mu U^{- 1} (y) \nonumber\\
  = &i g^{- 1} U (y) \partial_\mu U^{- 1} (y) \left(1 - \frac{\lambda^2}{y^2 + \lambda^2}\right) \nonumber\\
  = &i g^{- 1} U (y) \partial_\mu U^{- 1} (y) \frac{y^2}{y^2 + \lambda^2},
  \label{eq:A_transformed}
\end{align}
where %in the second equality 
we have used $\partial_\mu U (y) U^{- 1} (y) =- U (y) \partial_\mu U^{- 1} (y)$.
%$\partial_\mu U (y) U^{- 1} (y) = \partial_\mu (U (y) U^{- 1} (y)) - U (y) \partial_\mu U^{- 1} (y)=- U (y) \partial_\mu U^{- 1} (y)$. 

Similarly to $\bar{\Sigma}_{\mu \nu}$, we introduce 
\begin{align}
\Sigma_{\mu \nu} = \eta_{k \mu \nu} \frac{\sigma_k}{2}, \quad 
  \eta_{k \mu \nu} = - i \eta_{k \nu \mu} =
  \begin{cases}
    \epsilon_{k \mu \nu} &(\mu , \nu = 1 , 2 , 3) \\
    \delta_{k \mu}
  \end{cases}
\end{align}
to denote 
\begin{align}
  U (x) \partial_j U^{- 1} (x) 
%&= \frac{x_4 I + i x_i \sigma_i}{|x|}\left(- \frac{i \sigma_j}{|x|} - \frac{x_j U^{- 1} (x)}{|x|^2}\right) \nonumber\\
%&= \frac{- i x_4 \sigma_j + x_i (\delta_{i j} I + i \epsilon_{i j k} \sigma_k)}{|x|^2} - \frac{x_j I}{|x|^2} \nonumber\\
%&= i \frac{- x_4 \sigma_j - \epsilon_{k j i} x_i \sigma_k}{|x|^2} \nonumber\\
%&
= - 2 i \Sigma_{j \nu} \frac{x_\nu}{|x|^2},
\quad 
%\end{align}
%and
%\begin{align}
  U (x) \partial_4 U^{- 1} (x) 
%&= \frac{x_4 I + i x_i \sigma_i}{|x|} \left(\frac{I}{|x|} - \frac{x_4 U^{- 1} (x)}{|x|^2}\right) \nonumber\\
%&= \frac{x_4 I + i x_i \sigma_i}{|x|^2} - \frac{x_4 I}{|x|^2} \nonumber\\
%&= i \frac{x_i \sigma_i}{|x|^2} \nonumber\\
%&= i \frac{\delta_{i \nu} x_\nu \sigma_i}{|x|^2} \nonumber\\
%&
= - 2 i \Sigma_{4 \nu} \frac{x_\nu}{|x|^2} .
\end{align}
Therefore, the gauge transformation removes the singularity at $x = a_s$, and \eqref{eq:A_transformed} is cast into the instanton in the non-singular gauge which reproduces one instanton of the BPS type:
\begin{align}
  \mathscr{A}_\mu^\prime(x) = 2 g^{- 1} \Sigma_{\mu \nu} \frac{y_\nu}{y^2 + \lambda^2} = 2 g^{- 1} \Sigma_{\mu \nu} \frac{(x - a_s)_\nu}{(x - a_s)^2 + \lambda^2} .
\end{align}

%%%%%%%%%%%%%%%%%%%%%%%%%%%%%%%%%%%%%%%%%%%%%%%%%%%%%%%%%%%%%%%%%%%%%%%%%%%%%%%%%%%%%%%%%%%%%%%%%%%%%%%%%%%%%%%%%%

In order to understand the compact nature of the above argument, we introduce the three-dimensional unit vector field $\bm{n}(x) \ (\bm{n}(x) \cdot \bm{n}(x) = 1)$ by 
\begin{align}
  \bm{n} (x) = (\sin \beta (x) \cos \alpha (x) , \sin \beta (x) \sin \alpha (x) , \cos \beta (x)) \ (0 \leq \alpha \leq 2 \pi , \ 0 \leq \beta \leq \pi ).
\end{align}
The element $U$ of $SU (2)$ is represented in terms of the three Euler angles $\alpha$,  $\beta$, and $\eta$ as 
\begin{align}
  U (x) =& \exp \left(i \frac{\eta (x)}{2} \bm{n} (x) \cdot \bm{\sigma}\right) \in SU (2) \simeq S^3 \ (0 \leq \eta \leq 2 \pi ) .
%\nonumber\\
%& 0 \leq \alpha \leq 2 \pi , \ 0 \leq \beta \leq \pi , \ 0 \leq \eta \leq 2 \pi .
\end{align}
The two angles $\alpha$ and $\beta$ are used to determine the direction of the vector $\bm{n}$, while the remaining angle $\eta$ specifies the rotation angle around $\bm{n}$. 

The map $S^3 (\subset \mathbb{R}^4) \rightarrow \ SU (2) \simeq S^3$
%$x^\mu = (x_1 , x_2 , x_3 , x_4)$に対し！Ex^\mu = \sqrt{x^2} \left(\sin \frac{\eta (x)}{2} \bm{n} (x) , \cos \frac{\eta (x)}{2} \right)$
can be represented by using the parameterization of the coordinates:
\begin{align}
  x^\mu = (x_1 , x_2 , x_3 , x_4) = \sqrt{x^2} \left(\sin \frac{\eta (x)}{2} \bm{n} (x) , \cos \frac{\eta (x)}{2} \right) .
\end{align}
Indeed, we can check
\begin{align}
  x^\mu x^\mu = x^2 \left(\sin^2 \frac{\eta (x)}{2} \bm{n} (x) \cdot \bm{n} (x) + \cos^2 \frac{\eta (x)}{2}\right) 
  = x^2 \left(\sin^2 \frac{\eta (x)}{2} + \cos^2 \frac{\eta (x)}{2}\right) = x^2 .
\end{align}
In fact, this parameterization $S^3$ in  $\mathbb{R}^4$ in terms of the Euler angles reproduces the previous expression for the gauge transformation $U$ of $SU (2)$:
\begin{align}
  &U (x) = \cos \frac{\eta (x)}{2} + i \sigma^A \sin \frac{\eta (x)}{2} n^A (x) 
%= \frac{x_4}{\sqrt{x^2}} + i \bm{\sigma} \cdot\frac{\bm{x}}{\sqrt{x^2}}
%= \frac{x^4 + i \sigma^A x_A}{\sqrt{x^2}} 
= \frac{x_4}{\sqrt{x^2}} + i \sigma_k  \frac{x_k}{\sqrt{x^2}} , 
\end{align}
under the identification: 
\begin{align}
  &\cos \frac{\eta (x)}{2} = \frac{x_4}{\sqrt{x^2}} , \ \sin \frac{\eta (x)}{2} n^A (x) = \frac{x^A}{\sqrt{x^2}} \ (A = 1 , 2 , 3) \Leftrightarrow \ \sin \frac{\eta (x)}{2} = \frac{|\bm{x}|}{\sqrt{x^2}} ,
\end{align}
which yields
\begin{align}
  \frac{\eta (x)}{2} = \arccos \frac{x_4}{\sqrt{x^2}} = \arcsin \frac{|\bm{x}|}{\sqrt{x^2}} .
\end{align}
%where 
%$
% \left(\frac{x_4}{\sqrt{x^2}}\right)^2 + \left(\frac{|\bm{x}|}{\sqrt{x^2}}\right)^2 = \frac{x_4^2 + x_A^2}{x^2} = 1
%$.

\section{Details of calculating the integral}

By introducing the shifted variables $X = x - a_s , \ Y = y - a_s$, the integral is rewritten into
\begin{align}
  & \int d^D X e^{i p (X - Y)} \frac{X_\nu}{|X|^2} (\delta_{\mu \lambda} \Box_D - \partial_\mu \partial_\lambda) \frac{\frac{\Gamma \left(\frac{D}{2} - 1\right)}{4 \pi^{D / 2}}}{(|X - Y|^2)^{\frac{D - 2}{2}}} \nonumber\\
  = &i e^{- i p Y} \frac{\partial}{\partial p_\nu} \int d^D X e^{i p X} \frac{1}{|X|^2} \Biggl(\delta_{\mu \lambda} \delta^D (X - Y) \nonumber\\
  &- \frac{\Gamma \left(\frac{D}{2} - 1\right) (D - 2)}{4 \pi^{D / 2}} \frac{\delta_{\mu \lambda}}{|X - Y|^D} + \frac{\Gamma \left(\frac{D}{2} - 1\right) (D - 2) D}{4 \pi^{D / 2}} \frac{(X - Y)_\mu (X - Y)_\lambda}{|X - Y|^{D + 2}}\Biggl) \nonumber\\
  = &i e^{- i p Y} \delta_{\mu \lambda} \frac{\partial}{\partial p_\nu} \int d^D X e^{i p X} \frac{1}{|X|^2} \delta^D (X - Y) \nonumber\\
  &- i e^{- i p Y} \frac{\Gamma \left(\frac{D}{2} - 1\right) (D - 2)}{4 \pi^{D / 2}} \delta_{\mu \lambda} \frac{\partial}{\partial p_\nu} \int d^D X e^{i p X} \frac{1}{|X|^2} \frac{1}{|X - Y|^D} \nonumber\\
  &- i \frac{\Gamma \left(\frac{D}{2} - 1\right) (D - 2) D}{4 \pi^{D / 2}} \frac{\partial}{\partial p_\mu} \frac{\partial}{\partial p_\lambda} e^{- i p Y} \frac{\partial}{\partial p_\nu} \int d^D X e^{i p X} \frac{1}{|X|^2} \frac{1}{|X - Y|^{D + 2}} .
  \label{B1}
\end{align}

In particular, we consider the case $Y = 0$.
To perform the integral, we introduce the polar coordinates $(r , \theta , \cdots)$ for  the $D$-dimensional Euclidean space $X$ such that $r=|X|$ and $\theta$ is the angle between $X$ and $p$ with an arbitrary but fixed direction. Then the integrands depend only on $r$ and $\theta$.  Therefore, for the integration measure, it is enough to use the form:
\begin{align}
\int d^D X =  C_D \int_0^\infty d r \ r^{D-1} \int_{- 1}^1 d (\cos \theta) \sin^{D - 3} \theta
, \
C_D = \frac{2\pi^{\frac{D - 1}{2}}}{\Gamma \left(\frac{D - 1}{2}\right)} .
 \end{align}
 Then we have
\begin{align}
  ({\rm \ref{B1}})|_{Y=0} = &i \delta_{\mu \lambda} \frac{\partial}{\partial p_\nu} \int d^D X \delta^D (X) e^{i p X} \frac{1}{|X|^2}  \nonumber\\
   &- i \delta_{\mu \lambda} \frac{\Gamma \left(\frac{D}{2} - 1\right) (D - 2)}{4 \pi^{D / 2}} \frac{\partial}{\partial p_\nu} \int d^D X e^{i p X} \frac{1}{|X|^{D + 2}} \nonumber\\
   &- i \frac{\Gamma \left(\frac{D}{2} - 1\right) (D - 2) D}{4 \pi^{D / 2}} \frac{\partial}{\partial p_\mu} \frac{\partial}{\partial p_\lambda} \frac{\partial}{\partial p_\nu} \int d^D X e^{i p X} \frac{1}{|X|^{D + 4}} \nonumber\\
   = &i \delta_{\mu \lambda} \frac{\partial}{\partial p_\nu} \int_0^\infty d r   \delta (r) \int_0^\pi d \theta  \delta (\theta) e^{i |p| r \cos \theta} \frac{1}{r^2}  \nonumber\\
   &- i C_D \frac{\Gamma \left(\frac{D}{2} - 1\right) (D - 2)}{2 \pi^{D / 2}} \delta_{\mu \lambda} \frac{\partial}{\partial p_\nu} \int_0^\infty d r \int_{- 1}^1 d (\cos \theta) \sin^{D - 3} \theta \frac{e^{i |p| r \cos \theta}}{r^3} \nonumber\\
   &- i C_D \frac{\Gamma \left(\frac{D}{2} - 1\right) (D - 2) D}{2 \pi^{D / 2}} \frac{\partial}{\partial p_\mu} \frac{\partial}{\partial p_\lambda} \frac{\partial}{\partial p_\nu} \int_0^\infty d r \int_{- 1}^1 d (\cos \theta) \sin^{D - 3} \theta \frac{e^{i |p| r \cos \theta}}{r^5} \nonumber\\
   = &i \delta_{\mu \lambda} \frac{\partial}{\partial p_\nu} |p|^2 \int_0^\infty  d r \delta (r) \ e^{i r}  \frac{1}{r^2} \nonumber\\
   &- i C_D \frac{\Gamma \left(\frac{D}{2} - 1\right) (D - 2)}{2 \pi^{D / 2}} \delta_{\mu \lambda} \frac{\partial}{\partial p_\nu} |p|^2 \int_0^\infty \frac{d r}{r^3} \int_{- 1}^1 d (\cos \theta) \sin^{D - 3} \theta  e^{i r \cos \theta}  \nonumber\\
   &- i C_D \frac{\Gamma \left(\frac{D}{2} - 1\right) (D - 2) D}{2 \pi^{D / 2}} \frac{\partial}{\partial p_\mu} \frac{\partial}{\partial p_\lambda} \frac{\partial}{\partial p_\nu} |p|^4 \int_0^\infty \frac{d r}{r^5} \int_{- 1}^1 d (\cos \theta) \sin^{D - 3} \theta  e^{i r \cos \theta}  \nonumber\\
   = &2 i \delta_{\mu \lambda} p_\nu \int_0^\infty d r \delta (r) \ e^{i r}  \frac{1}{r^2} \nonumber\\
   &- i C_D \frac{\Gamma \left(\frac{D}{2} - 1\right) (D - 2)}{\pi^{D / 2}} \delta_{\mu \lambda} p_\nu \int_0^\infty \frac{d r}{r^3} \int_{- 1}^1 d (\cos \theta) \sin^{D - 3} \theta  e^{i r \cos \theta}  \nonumber\\
   &- 4 i C_D \frac{\Gamma \left(\frac{D}{2} - 1\right) (D - 2) D}{\pi^{D / 2}} (\delta_{\mu \lambda} p_\nu + \delta_{\nu \lambda} p_\mu + \delta_{\mu \nu} p_\lambda) \nonumber\\
   &\times \int_0^\infty \frac{d r}{r^5} \int_{- 1}^1 d (\cos \theta) \sin^{D - 3} \theta  e^{i r \cos \theta} ,
  \label{eq:condition_D}
\end{align}
where we have rescaled $r$ as $r \rightarrow \ r/|p| $ to factor out the $p$ dependence from the integrand.
The integrals are linear in $p_\nu, p_\mu, p_\lambda$ and vanish in the limit $p \to 0$.
This linear dependence in $p$ is guessed by the dimensional analysis from the beginning, because the other variables become dimensionless by this scaling. 
For more singular $\partial_\nu \omega(x)$, the power of $p$ could increase  and then go to zero more rapidly in the limit $p \to 0$. 

Here the singularity at $r=0$ namely at $x=a$  is avoided by introducing the short-distance cutoff $\epsilon > 0$ to make the integrals finite:
\begin{align}
  &\int_\epsilon^\infty d r \delta (r) \ e^{i r}  \frac{1}{r^2} \label{eq:condition_D_1} , \\
  &\int_{\epsilon}^\infty \frac{d r}{r^n} \int_0^\pi d \theta \sin^{D - 2} \theta \ e^{i r \cos \theta}  \ (n=3,5) . \label{eq:condition_D1} 
%  , \\
%&\int_{\epsilon}^\infty \frac{d r}{r^5} \int_0^\pi d \theta \sin^{D - 2} \theta \ e^{i r \cos \theta} 
%\int_{- 1}^1 d (\cos \theta) \sin^{D - 3} \theta   \label{eq:condition_D_3} ,
\end{align}

The integrals are  finite in the presence of the short-distance cutoff $\epsilon$, while it is divergent if $r=|X| = 0$ is included in the integration region. 
To see the finiteness of the integral, we can use 
$|\sin^{D - 2} \theta| \leq 1$ and $|e^{i r \cos \theta}| = 1$ to obtain the upper bound:
\begin{align}
 \left|  \int_{\epsilon}^\infty \frac{d r}{r^n} \int_0^\pi d \theta \sin^{D - 2} \theta  e^{i r \cos \theta} \right| \leq &  \int_{\epsilon}^\infty \frac{d r}{r^n} \int_0^\pi d \theta 
  =   \left[- \frac{\pi}{(n-1)r^{n-1}}\right]_\epsilon^\infty  
  =  \frac{\pi}{(n-1)\epsilon^{n-1}},%\\
%\left|\int_{\epsilon}^\infty d r \int_0^\pi d \theta \sin^{D - 2} \theta \frac{e^{i r \cos \theta}}{r^5}\right| \leq & \pi \int_{\epsilon}^\infty \frac{d r}{r^5}
\label{eq:condition_D_squeeze_1}
\end{align}
which is finite for $\epsilon > 0$.%, diverges $\epsilon \rightarrow \ 0$.

The angular integral can be done exactly  for any $D$ using the Bessel function $J_\nu (r)$. 

For $D=2$, 
\begin{align}
\int_{\epsilon}^\infty \frac{d r}{r^n} \int_0^\pi d \theta   \ e^{i r \cos \theta} 
=  \pi \int_{\epsilon}^\infty d r \frac{J_0(r)}{r^{n}} , 
\label{eq:condition_D_2} 
\end{align}
where the small $r$ expansion is given by
$
J_0(r) = 1 - \frac14 r^2 + O(r^4). 
$

For $D=4$, 
\begin{align}
\int_{\epsilon}^\infty \frac{d r}{r^n} \int_0^\pi d \theta \sin^{2} \theta  \ e^{i r \cos \theta} 
=  \pi \int_{\epsilon}^\infty  d r \frac{J_1(r)}{r^{n+1}}, 
\label{eq:condition_D_4} 
\end{align}
where the small $r$ expansion is given by
$
J_1(r) = \frac12 r [1 - \frac18 r^2 + O(r^4)]. 
$
The remaining $r$ integral can be done using the Gauss hypergeometric function, although we do not give the explicit form. 

For $D=3$,  in particular, the integration has the simple form:
\begin{align}
 \int_{\epsilon}^\infty \frac{dr}{r^n} \int_{- 1}^1 d (\cos \theta)  e^{i r \cos \theta} 
  = \int_{\epsilon}^\infty \frac{dr}{r^n}  \left[\frac{e^{i r \cos \theta}}{i r}\right]_{\cos \theta = - 1}^1
  =  \int_{\epsilon}^\infty d r \frac{2 \sin r}{r^{n+1}} .
\end{align}
The remaining integral is expressed by using the cosine integral. For instance, for $n=3$, by repeating integration by parts, we have
\begin{align}
  \int_{\epsilon}^\infty d r \frac{2 \sin r}{r^4} =  \frac{\text{Ci} (\epsilon)}{3} + \left[\frac{(r^2 - 2) \sin r - \cos r}{3 r^3}\right]_{\epsilon}^\infty
  =  \frac{\text{Ci} (\epsilon)}{3} - \frac{(\epsilon^2 - 2) \sin \epsilon - \cos \epsilon}{3 \epsilon^3} ,
  \label{eq:condition_D=3_2_int}
\end{align}
where we have introduced cosine integral
\begin{align}
  \text{Ci} (x) = - \int_x^\infty d t \frac{\cos t}{t} .
\end{align}
Anyway, these results agree with the bounds obtained in the above.

\end{document}